\documentclass[letterpaper,british,reprint, aps]{revtex4-1}
\usepackage[LGR,T1]{fontenc}
\usepackage[latin9]{inputenc}
\usepackage{array}
\usepackage{textcomp}
\usepackage{multirow}
\usepackage{amsmath}
\usepackage{amssymb}
\usepackage{graphicx}

\makeatletter

\pdfpageheight\paperheight
\pdfpagewidth\paperwidth

\DeclareRobustCommand{\greektext}{%
  \fontencoding{LGR}\selectfont\def\encodingdefault{LGR}}
\DeclareRobustCommand{\textgreek}[1]{\leavevmode{\greektext #1}}
\ProvideTextCommand{\~}{LGR}[1]{\char126#1}

\providecommand{\tabularnewline}{\\}

\makeatother

\usepackage{babel}
\begin{document}
\title{Investigating Top-Higgs FCNC Couplings at the FCC-hh }
\author{O. M. Ozsimsek}
\email{ozgunozsimsek@hacettepe.edu.tr}

\affiliation{Graduate School of Science and Engineering, Hacettepe University,
06800 Ankara, Turkey}
\author{V. Ari}
\email{vari@science.ankara.edu.tr}

\affiliation{Department of Physics, Ankara University, 06100 Ankara, Turkey}
\author{O. Cakir}
\email{ocakir@science.ankara.edu.tr}

\affiliation{Department of Physics, Ankara University, 06100 Ankara, Turkey}
\date{\today}
\begin{abstract}
We have studied the sensitivity to flavor changing neutral current
interaction of top and Higgs boson at the future circular collider
in the hadron-hadron collision mode (FCC-hh). Our main concerns are
the processes of $pp\rightarrow th$ (FCNC production) and $pp\rightarrow t\bar{t}$
(one top FCNC decay) which contributes to the single lepton, at least
three jets and the missing energy transverse in the final state. On
the one hand FCC-hh offers very high luminosity and large cross section
for these signal processes, on the other hand signal can be distinguished
from background which needs application of ingenious methods. Here,
we have inspired and followed the searches at the LHC using our analysis
path and enhanced attainable limits obtained on the possible top-Higgs
FCNC couplings phenomenologically. We obtain new limits which are
beyond the current experimental limits obtained from different channels
and processes at the LHC. The potential discovery or exclusion limits
on branching ratios for $tqH$ FCNC interactions can be set $BR(\eta_{uc})_{\mathrm{disc}}=9.08\times10^{-6}$
or $BR(\eta_{uc})_{\mathrm{exc}}=2.78\times10^{-6}$ at an integrated
luminosity of $30${\normalsize{} ab$^{-1}$} , respectively. Our
results are compatible with the other channels already studied at
FCC-hh.

\emph{Our work is presented in Arxiv with preprint number: arXiv:2204.13139 }
\end{abstract}
\pacs{1234}
\maketitle

\section{Introduction}

After the discovery of the top quark at the Tevatron experiments \citep{key-1-1,key-1-2},
and more recently the Higgs boson at the LHC experiments \citep{key-1-3,key-1-4},
it has been opened a new era of the searches for new physics beyond
the standard model (BSM). Conceptually the Standard Model (SM) suffers
from the naturalness/hierarchy problem \citep{key-4}. The problem
originates mainly from the interactions between Higgs and relatively
massive particles of SM which contributes the quantum corrections
of Higgs mass. As a consequence there is a hierarchy between energy
scales which gives known infinities, so there must be a new physics
at a higher energy regime to neutralize these infinities and stabilize
the electroweak (EW) scale.

In that sense, Higgs boson and top quark are the two heavy particles
at SM which implies the most sensitive to TeV scale physics. Besides
the contribution from loops with top quark to Higgs mass far more
dominant compared to others. Thus fate of BSM physics will be decided
by Higgs physics and top physics researches in that regard. 

In the quark sector, the flavor changing natural current (FCNC) interactions
\citep{key-55} are highly suppressed due to presence of CKM matrix.
New symmetries such as SUSY (supersymmetry) \citep{key-56-1,key-56-2,key-56-3,key-56-4}
or 2HDMs (two Higgs doublet models) \citep{key-57-1,key-57-2,key-57-3,key-57-4})
at BSM gives opportunities for new interacions. Seemingly there is
no reason to keep FCNC preserving structure of SM as we concern recent
findings at neutrino searches which allows mixings. Especially, some
models which will be mentioned at the following paragraphs have specifically
important for our research. But let us give some details before continuing:
FCNC interactions of MSSM (minimal supersymmetric standard model)
models predict a branching ratio $\leq\mathcal{O}(10^{-5})$ \citep{key-7-1,key-7-2,key-7-3},
same branchings are valid for composite Higgs models \citep{key-7-4}.
Additionally, 2HDMs foresee a brancing ratio roughly between $\mathcal{O}(10^{-3})$
to $\mathcal{O}(10^{-6})$ \citep{key-7-1,key-7-2,key-7-3}. Finally,
RS (Randall-Sundrum) model and quark singlet models are expected to
have a branching ratio about $\mathcal{O}(10^{-4})$ \citep{key-7-1,key-7-2,key-7-3}. 

Many experiments have been performed to find some evidence for FCNC
interactions. However, the limits on FCNC couplings and branching
ratios have been set by ATLAS and CMS collabarations. Upper limits
for branching ratio by the ATLAS collaboration for the $BR(t\to cH)$
and the $BR(t\to uH)$ observed (expected) are $1.1\times10^{-3}$
$(8.3\times10^{-4})$ and $1.2\times10^{-3}$ $(8.3\times10^{-4})$
at the $95\%$ confidence level respectively \citep{key-6-1}. In
this paper, coupling constants for $tcH$ and $tuH$ limited observed
(expected) as 0.064 (0.055) and 0.066 (0.055), respectively \citep{key-6-1}.
Similar study by the CMS excludes limits as $BR(t\to uH)<1.9\times10^{-4}$
$(3.1\times10^{-4})$ and $BR(t\to cH)<9.4\times10^{-4}$ at $95\%$
confidence level respectively for $H\to\gamma\gamma$ channel \citep{key-6-2}.
Here coupling constants are given as upper limits with 95\% CL, observed
(expected) for $tcH$ and $tuH$ 0.071 (0.060) and 0.037 (0.047) ,
respectively \citep{key-6-2}. Additionaly, for $H\to b\bar{b}$ channel
similar research set limits for branchings observed (expected) such
that $BR(t\to uH)<7.9\times10^{-4}$ $(1.1\times10^{-3})$ and $BR(t\to cH)<9.4\times10^{-4}$
$(8.6\times10^{-4})$ by the CMS \citep{key-6-3}. At this research
limits on couplings also determined as observed (expected) for $tcH$
and $tuH$ 0.081 (0.078) and 0.074 (0.087) respectively at 95\% CL
\citep{key-6-3}. This research is followed by a new study by the
ATLAS which gives limits as $BR(t\to uH)<7.2\times10^{-4}$ $(3.6\times10^{-3})$
and $BR(t\to cH)<9.9\times10^{-4}$ $(5.0\times10^{-4})$ respectively
\citep{key-6-4}. Then experimental limits are at the edge of $\mathcal{O}(10^{-4})$
at best for branchings and the coupling constants are expected to
be below roughly 0.05. 
\begin{figure}[h]
\begin{raggedright}
\includegraphics[scale=0.48]{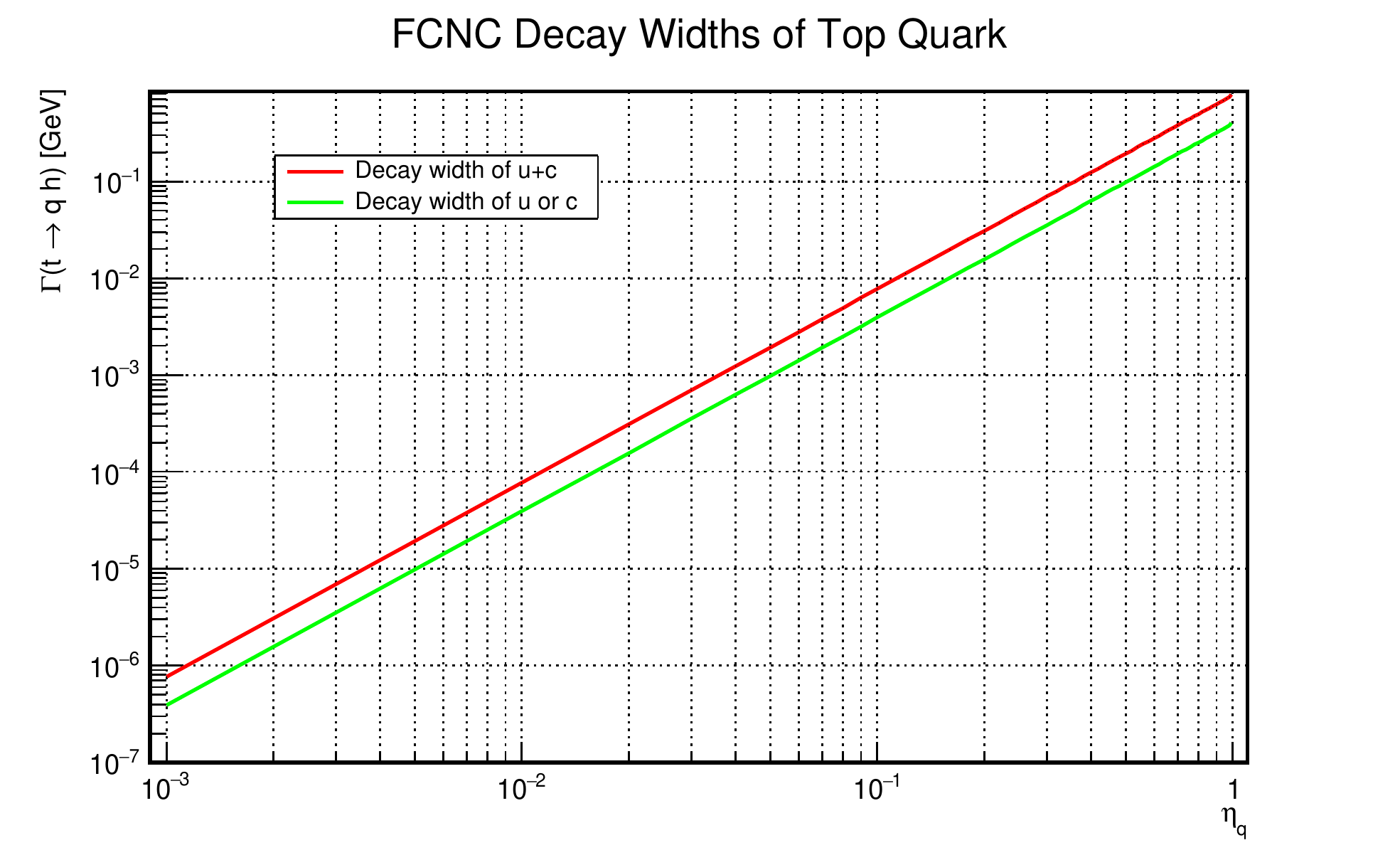}
\par\end{raggedright}
\raggedright{}\caption{The FCNC decay width of top quark according to two scenarios: For
$u+c$ case top quark can decay into both, otherwise decays into only
one of them.}
\end{figure}
\begin{figure}[h]
\begin{raggedright}
\includegraphics[scale=0.48]{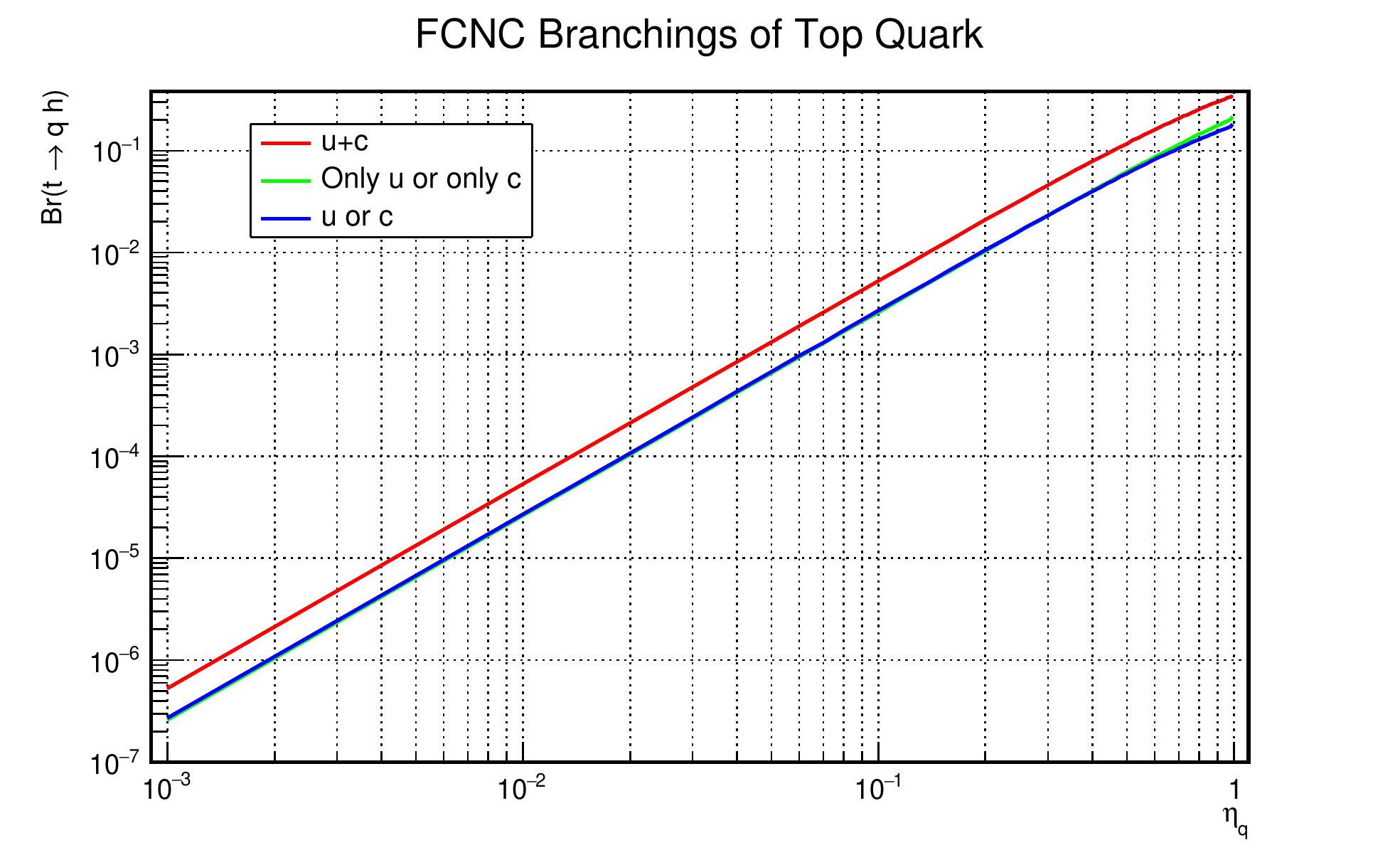}
\par\end{raggedright}
\raggedright{}\caption{FCNC branchings of top quark for three scenario presented: $u+c$
indicates top quark can decay both $u$ and $c$. $u$ or $c$ is
similar but only one channel preferred. Hence branchings instantly
declined to approximately to half. Only $u$ or only $c$ indicates
only one of the decay channel is accessible. }
\end{figure}

Future colliders having higher luminosity and higher center of mass
energy are on the agenda. In addition to this, conceptual desing report
(CDR) of FCC-hh have been also published \citep{key-7-2}. Especially
with 100 TeV center of mass energy and 30 $ab^{-1}$ luminosity FCC-hh
enables for discovering or excluding many BSM scenarios which excites
whole physics community. It is obvious that FCC-hh will enhance limits
for all type of interactions. Bear in mind top quark and Higgs boson
has vital importance for new physics, so they have the leading role
also in FCNC interactions. In this study we would like to contribute
expectations for possible FCC-hh discovery and or exclution limits
for Higgs-top quark FCNC interactions. While finishing, to close the
circle by stressing the iportance of FCC we would like to say the
expected branchings for FCNC interactions at FCC-hh are about $\mathcal{O}(10^{-5})$;
we are looking for limits comparable to these expectations. It is
obvious that with this capacity FCC-hh can rule out some RS, 2HDMs
(especially flavor violating) and quark singlet models. Moreover,
taking advantage of its high COM and luminosity, it can also penetrate
the MSSM and some other 2HDMs regions where FCNC interactions are
predicted. So, we take into account the process $pp\rightarrow th(q)\quad,(t\rightarrow W^{+}+b,\ W^{+}\rightarrow l^{+}+\nu_{l^{+}}),\ (h\rightarrow b\bar{b})$
interactions (top quark and Higgs produced on-shell) at the FCC-hh
and try to exploit the potential discovery and exclusion limits of
FCNC interactions. 

\section{Model Framework}

FCNC interactions studied at numerous scenarios depending on models
which have many simple/complex extentions over the SM. Nevertheless,
we would like to investigate the problem in the context of EFT, which
is quite flexible and suitable for comparing results with experimental
results. In that respect EFT provides a model independent research
method to exploit the potentials of the new colliders. 

FCNC interactions including a $tqH$ vertex in generic manner can
be described using an effective lagrangian as given as 

\begin{align}
L_{H} & =\frac{1}{\sqrt{2}}H\bar{t}(\eta_{u}^{L}P^{L}+\eta_{u}^{R}P^{R})u+h.c.\nonumber \\
 & +\frac{1}{\sqrt{2}}H\bar{t}(\eta_{c}^{L}P^{L}+\eta_{c}^{R}P^{R})c+h.c.\label{eq:1}
\end{align}
where the $\eta_{q}^{L/R}$ couplings shows the strenght of interactions
among $u$ or $c$ to top quark and Higgs. Superindices $L,R$ denotes
chirality which can be taken equal. 

We can define more couplings as imaginary numbers but for sake of
simplycity we omit those and pick couplings as real numbers. However,
it should be noted that we let the number $\frac{1}{\sqrt{2}}$ in
Lagrangian to restrict the discovery/exclusion region. It is obvious
that it effects both branching ratios and cross sections. 

By using that Lagrangian we calculated decay width as

\begin{equation}
\Gamma(t\to qh)=\frac{(\eta_{qL}^{2}+\eta_{qR}^{2})}{64\pi}\frac{(m_{t}^{2}-m_{h}^{2})^{2}}{m_{t}^{3}}
\end{equation}
After we insert all parameters we find depency of decay width depending
on FCNC couplings 
\begin{equation}
\Gamma(t\to qh)\simeq0.1904(\eta_{qL}^{2}+\eta_{qR}^{2})\ \mathrm{GeV}
\end{equation}
From the general definition of the branching ratio to FCNC interactions,
relation turns into 
\begin{equation}
BR(t\to qh)=\frac{\Gamma(t\to qh)}{\Gamma(t\to qh)+\Gamma(t\to Wq)}
\end{equation}
Decay of top quark is special: it decays before hadronisation and
leaves a trace with asymmetrical spin behaviour. In the case of decay
modes which are dominated by $t\to Wb$ procces, thus leads quadratic
increment in branching ratio for small couplings as other contributions
will be neglected in denominator. 

It is already known that FCNC interactions are suppressed in one loop
in SM. In EFT we use dimension six operators as the source of these
interactions. To be specific about the model and operators we have
introduced, the EFT Lagrangian
\begin{equation}
\mathcal{L}^{\mathrm{eff}}=\sum\frac{C_{x}}{\Lambda^{2}}O_{x}+...
\end{equation}
where $C_{x}$ are being complex constants and $O_{x}$ are related
dimension-six gauge-invariant operators \citep{key-3-1,key-3-2}.
High powers of $\Lambda$ will suppress other terms and we focus mainly
on the contributions comes from the first term. Relevant operators
are $(i\ne j)$
\begin{equation}
\frac{i}{2}\left[\phi^{\dagger}\tau^{I}D_{\mu}\phi-\left(D_{\mu}\phi\right)^{\dagger}\tau^{I}\phi\right]\left(\bar{q}_{L_{i}}\gamma^{\mu}\tau^{I}q_{L_{j}}\right)
\end{equation}
\begin{equation}
\frac{i}{2}\left(\phi^{\dagger}\overleftrightarrow{D}_{\mu}\phi\right)\left(\bar{q}_{L_{i}}\gamma^{\mu}q_{L_{j}}\right)
\end{equation}
\begin{equation}
\frac{i}{2}\left(\phi^{\dagger}\overleftrightarrow{D}_{\mu}\phi\right)\left(\bar{u}_{R_{i}}\gamma^{\mu}u_{R_{j}}\right)
\end{equation}
where $i,j$ indices show flavor. $\tau^{I}$ shows Pauli matrices
$R,L$ indices shows chiral states as usual. For a complete picture
let us show the relationship between both frameworks over coefficients
\begin{equation}
\delta\eta_{ct}^{L}=-\frac{3}{2}C_{u\phi}^{32*}\frac{v^{2}}{\Lambda^{2}}\quad,\quad\delta\eta_{ct}^{R}=-\frac{3}{2}C_{u\phi}^{23}\frac{v^{2}}{\Lambda^{2}}
\end{equation}

\begin{equation}
\delta\eta_{ut}^{L}=-\frac{3}{2}C_{u\phi}^{31*}\frac{v^{2}}{\Lambda^{2}}\quad,\quad\delta\eta_{ut}^{R}=-\frac{3}{2}C_{u\phi}^{13}\frac{v^{2}}{\Lambda^{2}}
\end{equation}

then we may correlate our result with the previous studies/analyses
\citep{key-3-1,key-3-2}.

\section{Cross Sections of Signal and Background}

After we introduce our framework we would like to present our signal
process, decay channel and possible backgrounds before analysis. 

\begin{figure}[h]
\includegraphics[scale=0.3]{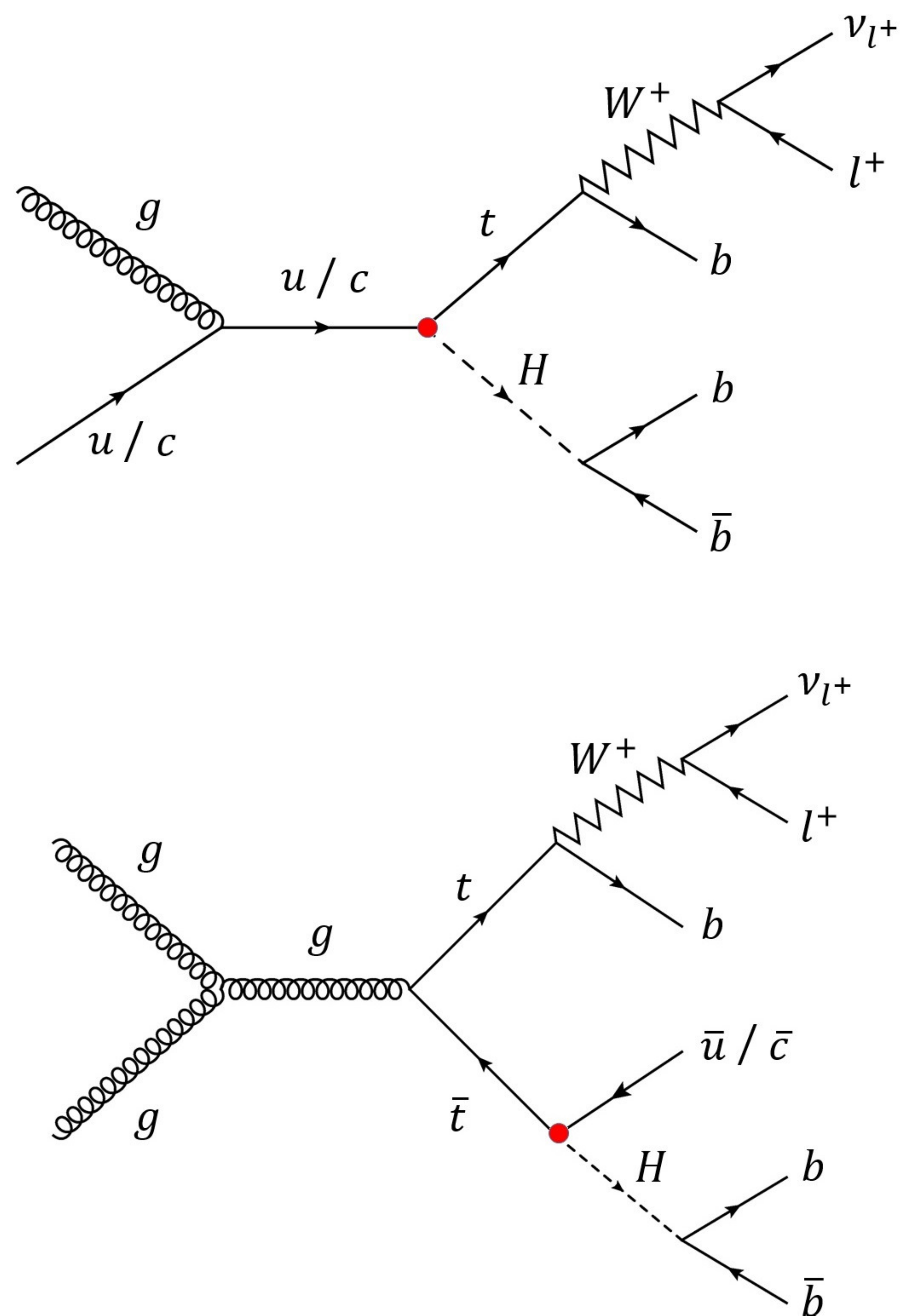}

\caption{Representative Feynman diagrams of production and decay channels\label{fig:Representative-Feynman-diagrams}}

\end{figure}

We choose both $pp\rightarrow t(\bar{t})h$ and $pp\rightarrow t(\bar{t})hj$
as signal processes (see Fig. \ref{fig:Representative-Feynman-diagrams})
in order to give Higgs boson and top quark to be products leading
to single lepton, missing transverse momentum and the multijet final
state. Here, we expect to enhance signal by taking them both of them
and include the processes with $\bar{t}$ to collect more events. 

Since, we would like to optimize cross section to observe maximal
number of events while keeping the problem realistic with relevant
backgrounds, we impose sufficient conditions at every step of decisions.
The signal $2\rightarrow2$ type process has high cross section as
we know this kind of interactions which gives better statistics. The
other process is $2\rightarrow3$ type, its cross section is also
high.

Our aim is to work with a hadron collider at 100 TeV which produces
many jets due to nature of interactions, thus at least one leptonic
channel would give clear signal footprint. Despite the leptonic channel
lowers the cross section, it offers a better analysis.

As we know new physics interactions at BSM introduces a new vertex
at effective Lagrangian with a constant absolute value in range {[}0,1).
Hence more new physics vertices give lower cross sections and make
analysis harder. Thus we would like to restrict ourselves with less
parameters if possible. Furthermore, we would like to avoid interferance
effects as possible, then we generate events having this property.

We use leptonic decay of $W$ boson to get rid of jet mess while reconstructing
top and Higgs, besides Higgs decays to the channel where branching
is the highest to get a higher number of events ($b,\bar{b}$). This
leads much more statistics if signal and and background can be analysed
properly, otherwise advantage of leptonic channel can be lost and
there can be still jet mess present. Note that a research for similar
channel has been done by CMS Collaboration at the LHC \citep{key-6-3},
we would like to enlarge scope by letting associated jets, and project
it to FCC-hh. In that analysis, analysis region divided so that jet
number restiricted to three or four and two or three of them tagged
as $b$; besides machine learning techniques used. Here, we are using
cut-based analysis and to deal with jet mixings we use different analyse
path while tagging b-jets which is crucial ingredient of the analysis.
For now, we would like to stress the importance of this issue and
postpone it to several paragraphs later.

We deal with three scenarios while approaching the problem and would
like to discuss results in that perspective in order to get situations
where at least one of the coupling constant different from zero. While
analyzing we will use tree scenario namely $u+c$, only $u$ and only
$c$. 
\begin{figure}[h]
\includegraphics[scale=0.48]{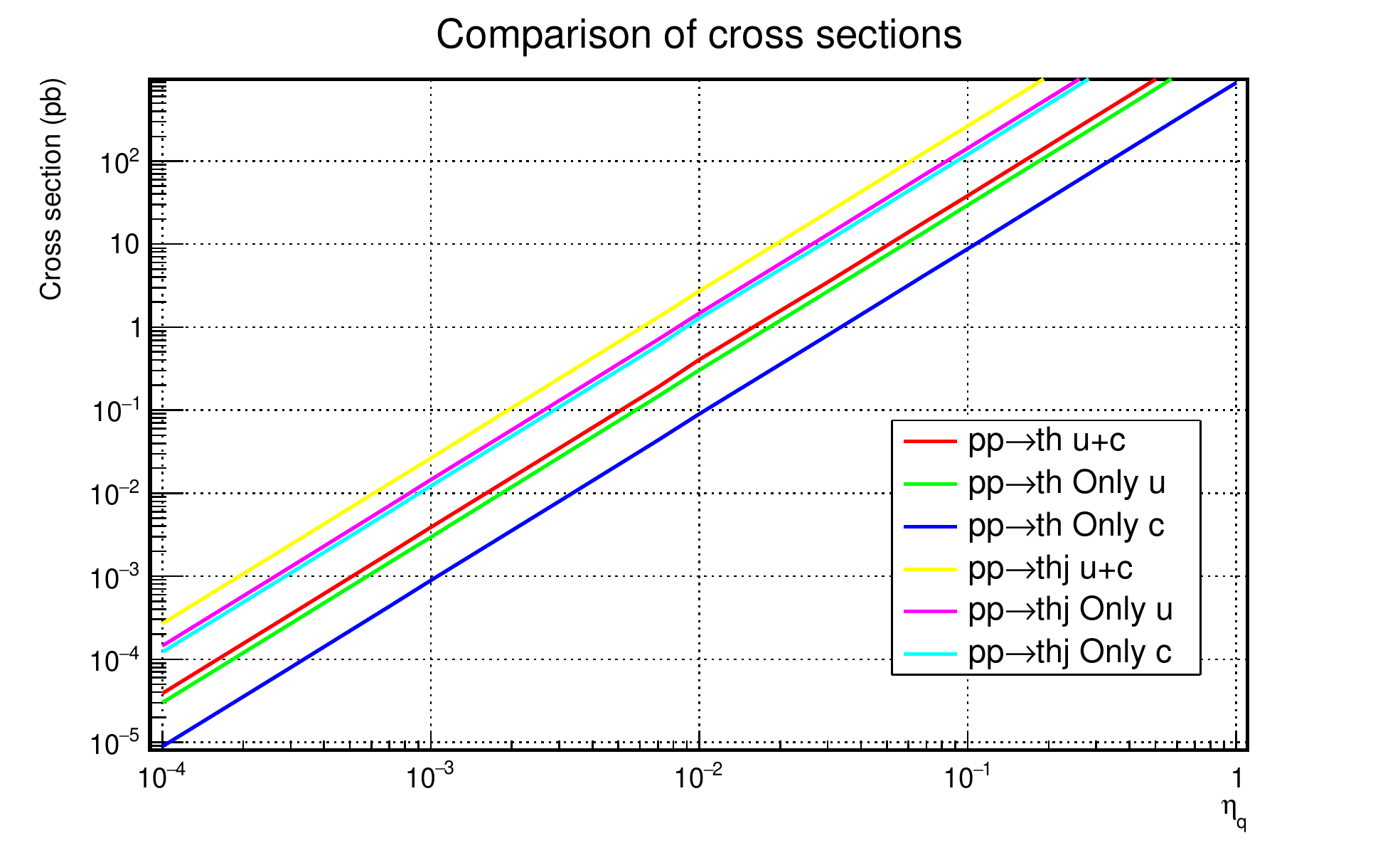}\caption{Comparison of FCNC cross sections: When two FCNC channels are open
the cross section is highest for two type of processes as expected.
Cross section of $pp\rightarrow thj$ process is also higher nearly
one order of magnitude. At hadron colliders there are numerous final
jets as a result of nature of interactions, and allowance one jet
near interested particles increases cross section considerably, even
it is a $2\rightarrow3$ process compared to the $2\rightarrow2$
process. Similar to other considerations argued before, there is an
optimisation, thus producing more jets is making tagging harder and
lowers the relevance of FCNC process.}

\end{figure}

The $u$ quark contribution is higher than the $d$ quark to the cross
section. This is realised so that up quarks are being valance quarks
in proton which have higher distributions. After initial setup, Madgraph5
\citep{key-4-1} is used to generate signal and background samples
by using parton distribution function (PDF) set NNPDF2.3 \citep{key-4-7}
at five flavor scheme (topFCNC model was used) \citep{key-4-2,key-4-3}.
Then by using PYTHIA8 \citep{key-4-4}, samples generated at partonic
level have been hadronised which is followed by fast detector simulation
with DELPHES 3 \citep{key-4-5} by using FCC-hh detector card. The
samples are produced with default cuts 
\begin{enumerate}
\item $p_{T}^{\mathrm{jets}}>20\ \mathrm{GeV},\ |\eta^{\mathrm{jets}}|<5$
\item $p_{T}^{\mathrm{leptons}}>10\ \mathrm{GeV},\ |\eta^{\mathrm{leptons}}|<2.5$
\item $\Delta R(i,j)>0.4,\ i,j$ being jets and leptons. 
\end{enumerate}
Finally Root6 \citep{key-4-6} is used to analyse resultant sample
files. Our signal final state has one lepton, missing transverse energy
and at least three jets (where at least two of them required as $b$
jets) as the characteristic.
\begin{table}[h]
\raggedright{}\caption{\label{tab:We-impose-(anti)top}We impose top or anti-top quarks to
decay into $Wb$ and $W$ to decay leptonically for each case for
maximal mixings. Higgs boson is allowed to decay to $b$ pair. After
we get one lepton and at least three jets we let other particles to
decay freely to get a realistic scenario with high cross sections
from processes. Moreover considering b-tag efficiencies we do not
allowed the other jets produced alongside with top quark and Higgs
to be as $b$ jets. The symbol $B$ indicates $Z$ and $W$.}
\begin{tabular}{|c|c|c|}
\hline 
Process & Cross section(pb) & Final states\tabularnewline
\hline 
\hline 
$pp\rightarrow t(\bar{t})h$ &  & $l^{\pm},\!\nu_{l^{\pm}},\!b(\bar{b}),\!2j$\tabularnewline
$pp\rightarrow t(\bar{t})hj$ &  & $l^{\pm},\!\nu_{l^{\pm}},\!b(\bar{b}),\!3j$\tabularnewline
$\eta_{u}=\eta_{c}=0.0075$ & $0.3766$ & \tabularnewline
$\eta_{u}=0.0075$ & $0.2203$ & \tabularnewline
$\eta_{c}=0.0075$ & $0.1536$ & \tabularnewline
\hline 
$pp\rightarrow t\bar{t}j$ & $1.576\times10^{4}$ & $l^{\pm},\!\nu_{l^{\pm}},\!2b(\bar{b}),\!j$\tabularnewline
\hline 
$pp\rightarrow t\bar{t}jj$ & $1.656\times10^{4}$ & $l^{\pm},\!\nu_{l^{\pm}},\!2b(\bar{b}),\!2j$\tabularnewline
\hline 
$pp\rightarrow t\bar{t}h$ & $15.33$ & $l^{\pm},\!\nu_{l^{\pm}},\!2b(\bar{b}),\!2j$\tabularnewline
\hline 
$pp\rightarrow t\bar{t}B$ & $14.18$ & $l^{\pm},\!\nu_{l^{\pm}},\!2b(\bar{b}),\!2j$\tabularnewline
\hline 
$pp\rightarrow thj$ & $0.6787$ & $l^{\pm},\!\nu_{l^{\pm}},\!b(\bar{b}),\!3j$\tabularnewline
\hline 
$pp\rightarrow Whjj$ & $2.75$ & $l^{\pm},\!\nu_{l^{\pm}},\!2b(\bar{b}),\!2j$\tabularnewline
\hline 
\end{tabular}
\end{table}

Here we would like to mention some other features of signal and background
and we will present our analysis strategy at the next section. We
observe two features at backgrounds: Top pair plus jet(s) have significantly
high cross section while extra particle content mainly includes more
jets which increases $H_{T}$ but three $b$ jets are not guaranteed.
The others containing Higgs or another boson, still do not guarantee
three $b$ jets and there are extra jets which also increases total
hadronic transverse energy $H_{T}$ while having low cross sections.
The $B$ (any vector boson) decays to jets, similarity with signal
is slightly higher at the backgrounds which have relatively low cross
sections. In the case of top pair plus jet(s), their large cross sections
would dominate whole region, hence it seems impossible to get rid
of them completely. 

An important aspect of these interactions are fat jets which are hitting
the nearly same region of the detector. Momentum of jets are also
crucial while analyzing, many of them comes from decay of two heaviest
particles within the SM. There can also be resudually produced jets.
Moreover, boosted objects with additional decay process including
top quark affects the reconstruction of Higgs, likewise extra top
quark which is not decaying leptonically gives alos a pseudo Higgs
reconstruction, then such backgrounds including top pair need evolved
eloboration.

Additionally, there are jets produced alongside with our mainly interested
particles and they are produced residually which should not be $b$
jets necessarily. Thus, some parts of our bacgrounds contains at least
two or three $b$ jets. Furthermore, mixings and tagging efficiencies
plays a major role here, one can not simply analyze signal by only
tagging the jets, detector effects such as misidentification of particles,
loosing particles or overcounting them require much evolved methods
to analyze while tagging $b$ jets there is another optimisation,
unique to analysis to eliminate background and still have reasonable
number of events. One can put forward several numbers of b-tagging
as a categorization when the analyse investigated. However, for a
cut based analysis at a hadron factory, tagging jets and putting kinematical
cuts is not a viable option to analysis (see Fig. \ref{fig:In-the-analysis,},
\ref{fig:In-the-analysis,-1}). We will come to this point at analysis
part again due to its importance while reconstructing objects.

To sum up, When all these effects combined with high cross section,
result is more realistic physically as we try to get. 
\begin{figure}[h]
\includegraphics[scale=0.45]{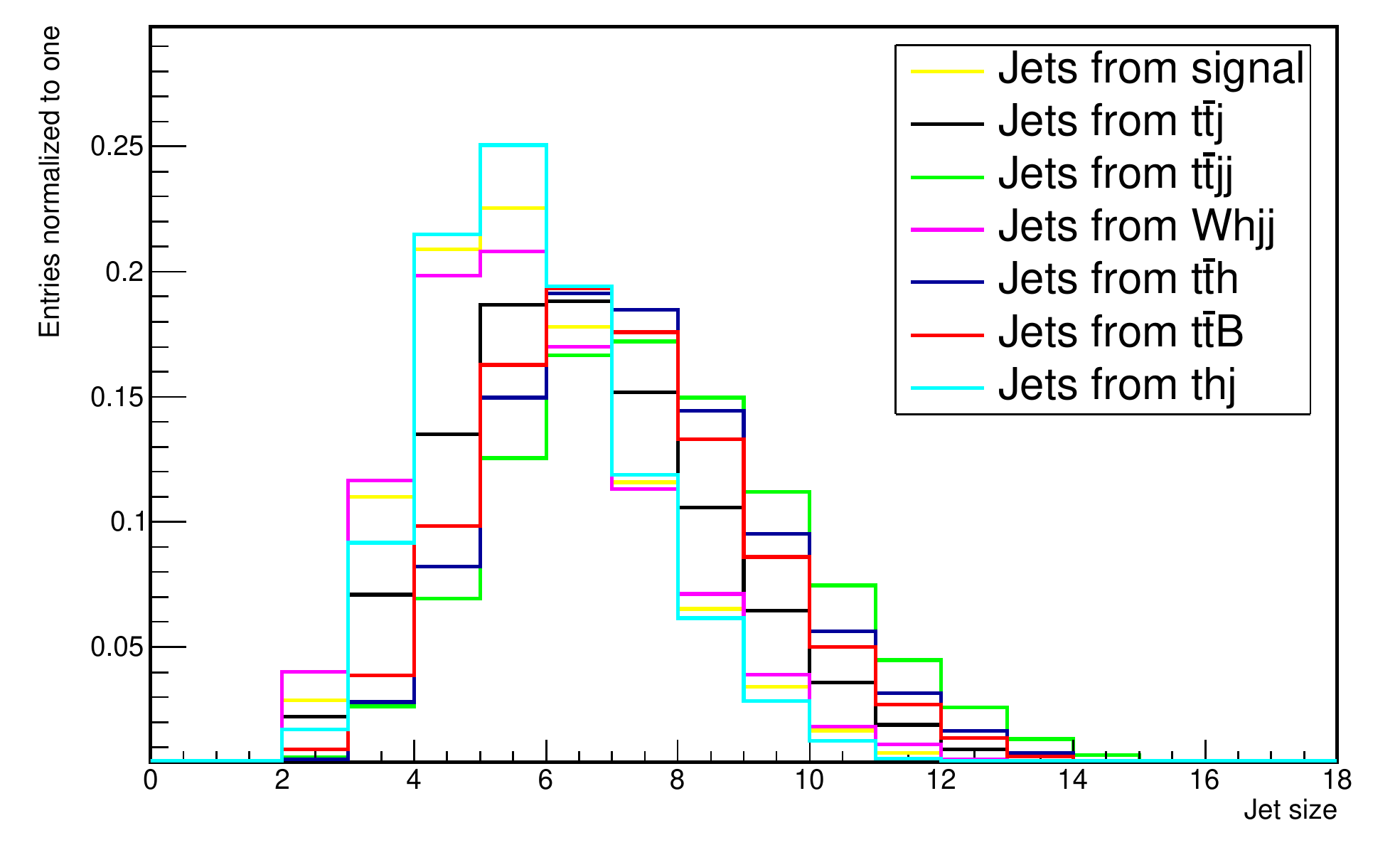}

\caption{In the analysis, at least three jets are selected, and used in reconstruction
according to their $p_{T}$ order. It is expected that after collision
many jets are produced both signal and background events. Observe
that, peak of the distribution lower at signal case, because of the
low number of original final state particles.\label{fig:In-the-analysis,}}
\end{figure}
\begin{figure}[h]
\includegraphics[scale=0.45]{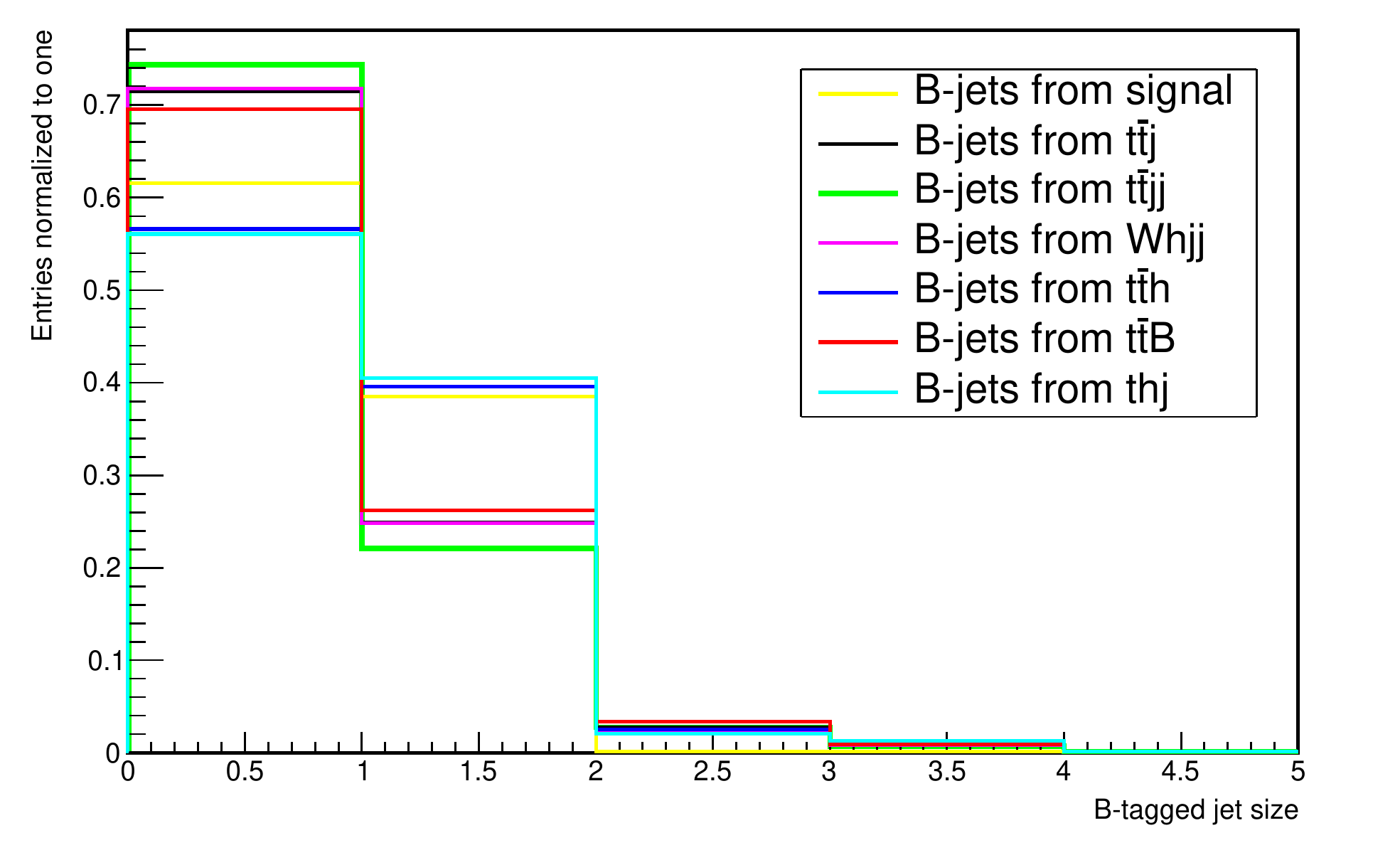}\caption{In the analysis, at least one b-tagged jet is selected with an efficiency
resulting from detector simulation. Number of events decreased drastically
if one tries to tag more jets. Tagging three jets as an analysis method
is not always a viable option.\label{fig:In-the-analysis,-1}}
\end{figure}

Before we advance the analysis we would like to explain briefly why
we let Higgs to decay into $b$ jets directly at process shown as
$pp\rightarrow Whjj$. It differs from other backgrounds not including
top quark which shifts signal reconstruction region drastically. It
has low cross section among other backgrounds. Hence, we would like
to aid and restore this background a little bit while increasing its
similarity with signal.

\section{Analysis}

\begin{figure}[h]
\raggedright{}\includegraphics[scale=0.46]{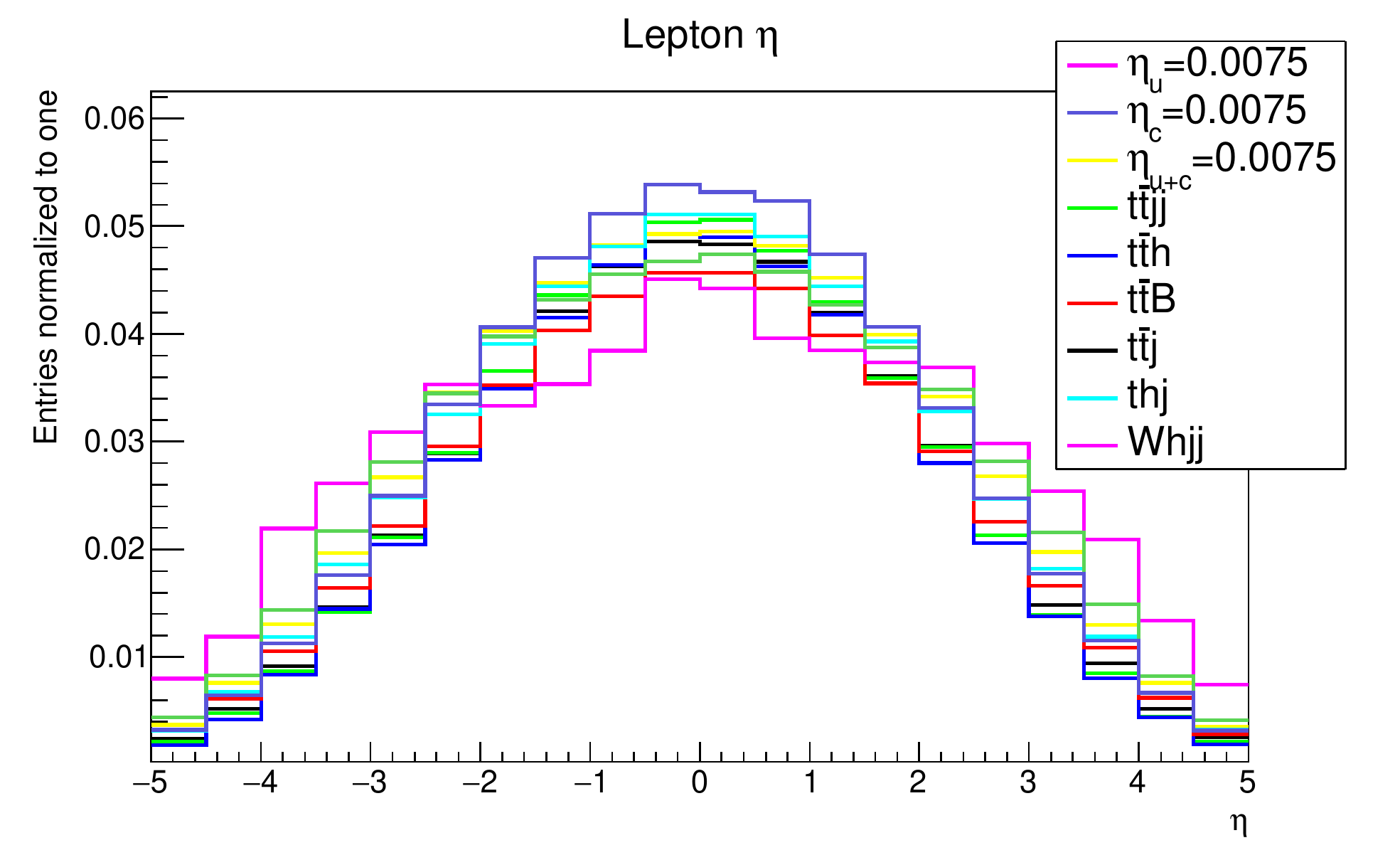}\caption{Lepton $\eta$ distributions showing general detection is central.\label{fig:Lepton--distributions}}
\end{figure}

\begin{figure}[h]
\includegraphics[scale=0.48]{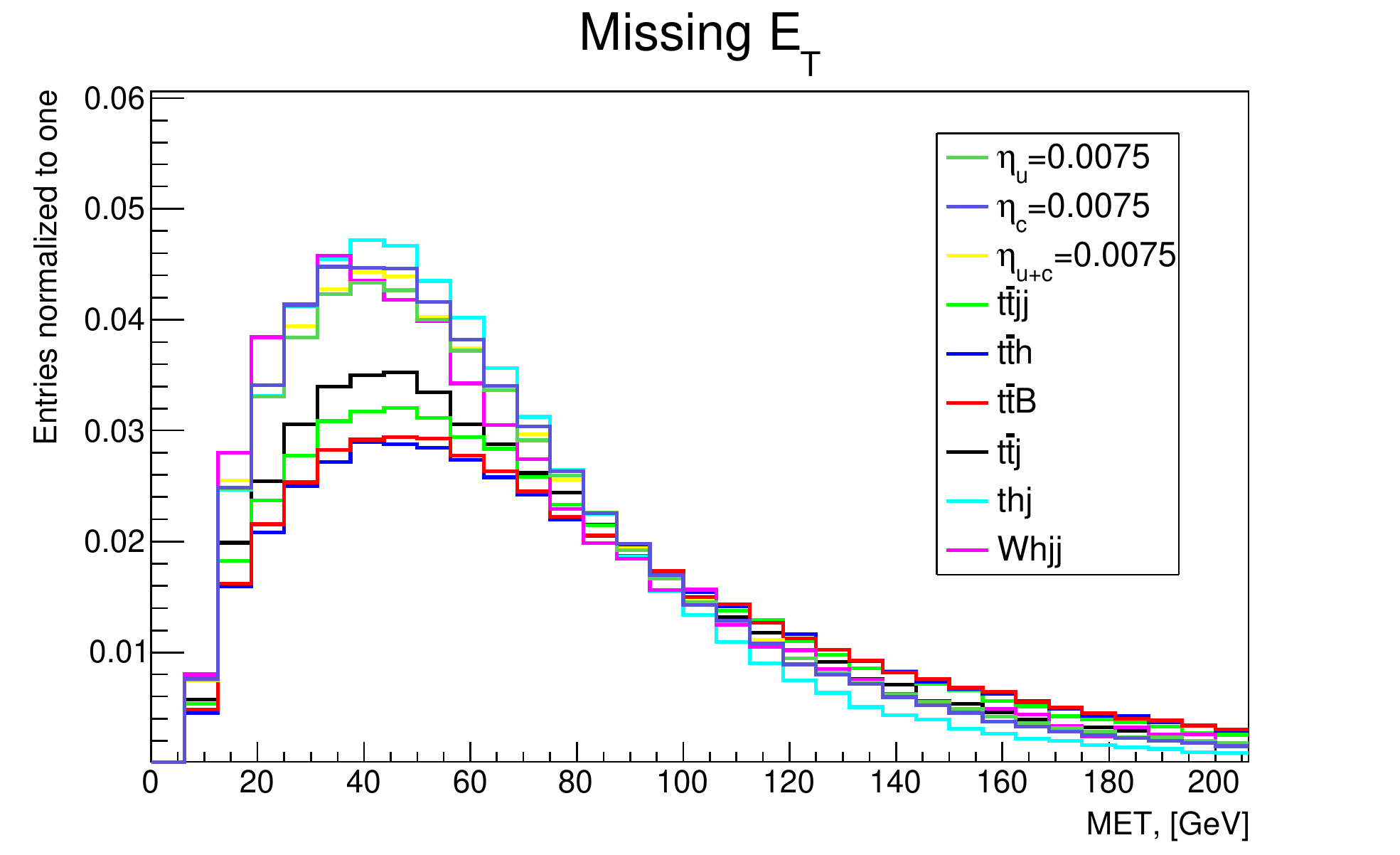}
\raggedright{}\caption{MET distribution for signal and relevant backgrounds.\label{fig:MET-distribution-for}}
\end{figure}

\begin{figure}[h]
\includegraphics[scale=0.48]{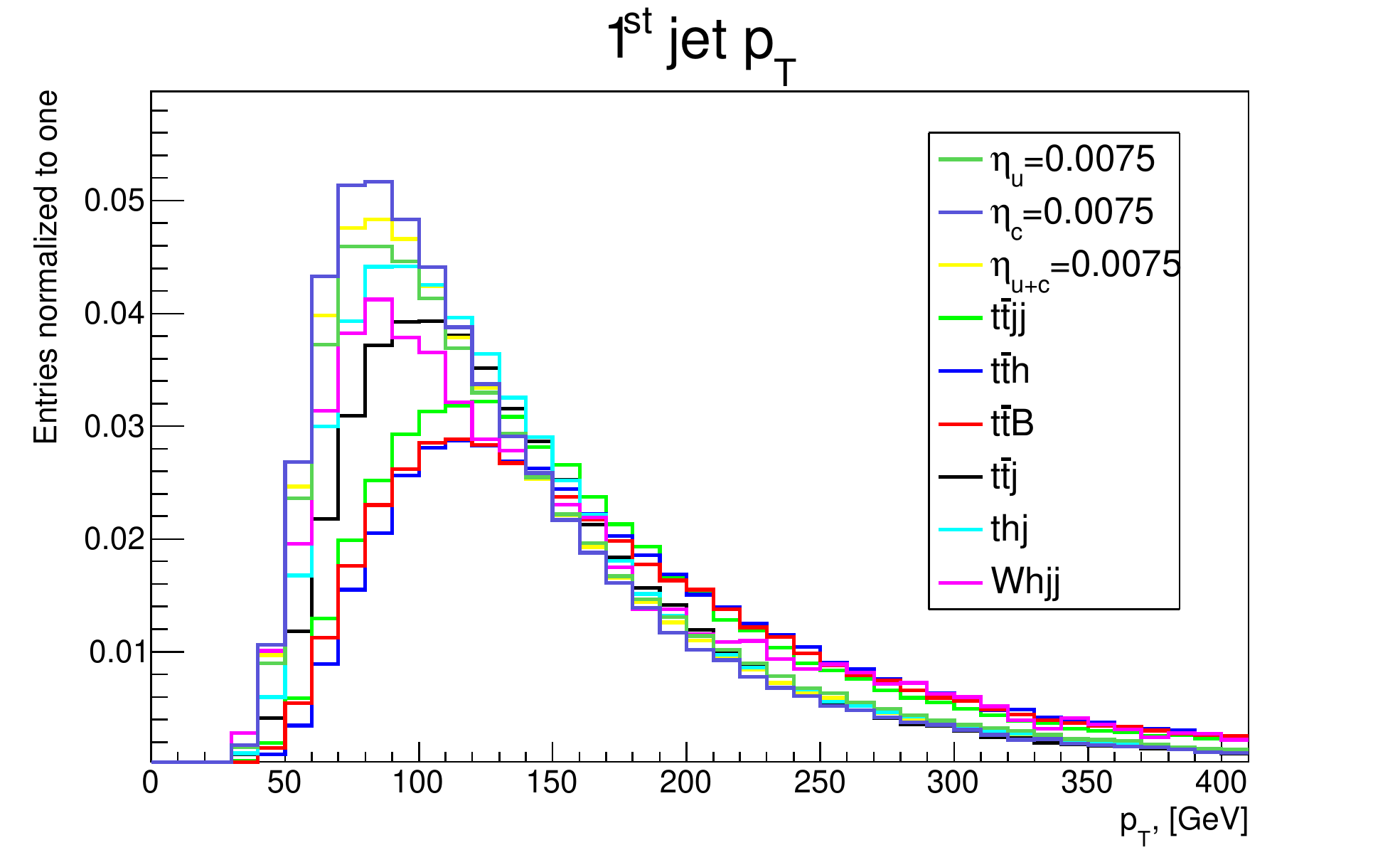}
\raggedright{}\caption{$1^{\mathrm{st}}$ jet $p_{T}$: Distributions each having high values.
More over they have a peak arround 80 GeV shows that they are mainly
coming from disintegration of top quark. Here, boosted structure of
jets can be seen.\label{fig:-jet-:pt1}}
\end{figure}

\begin{figure}[h]
\raggedright{}\includegraphics[scale=0.48]{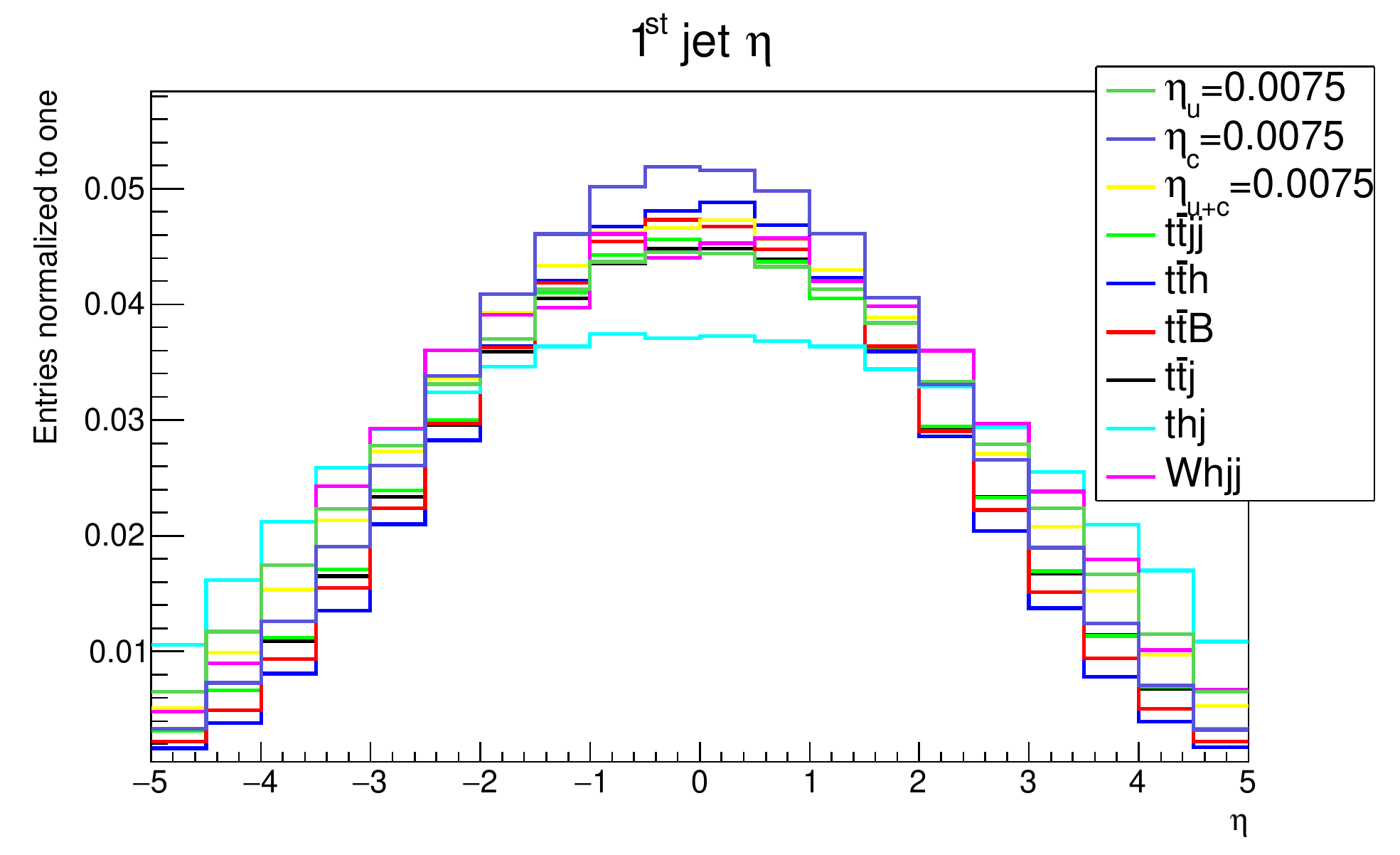}\caption{$1^{\mathrm{st}}$ jet $\eta$ distributions at detector which are
mostly central.\label{fig:-jet-eta1}}
\end{figure}

\begin{figure}[h]
\includegraphics[scale=0.48]{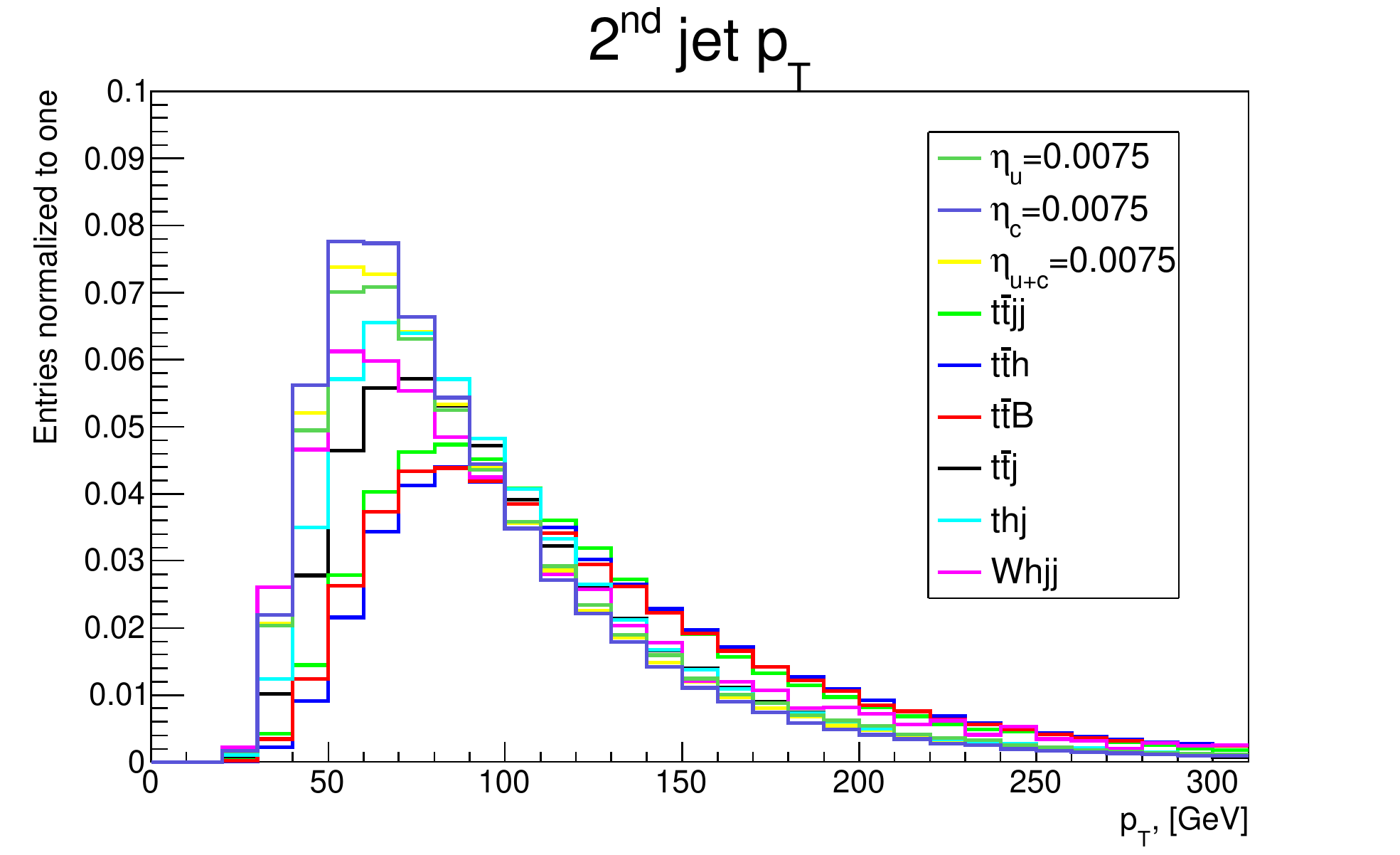}

\caption{$2^{\mathrm{nd}}$ jet $p_{T}$: Distributions with peak arround 60
GeV shows they are elements of Higgs decay.\label{fig:-jet-:pt2}}
\end{figure}

\begin{figure}[h]
\includegraphics[scale=0.48]{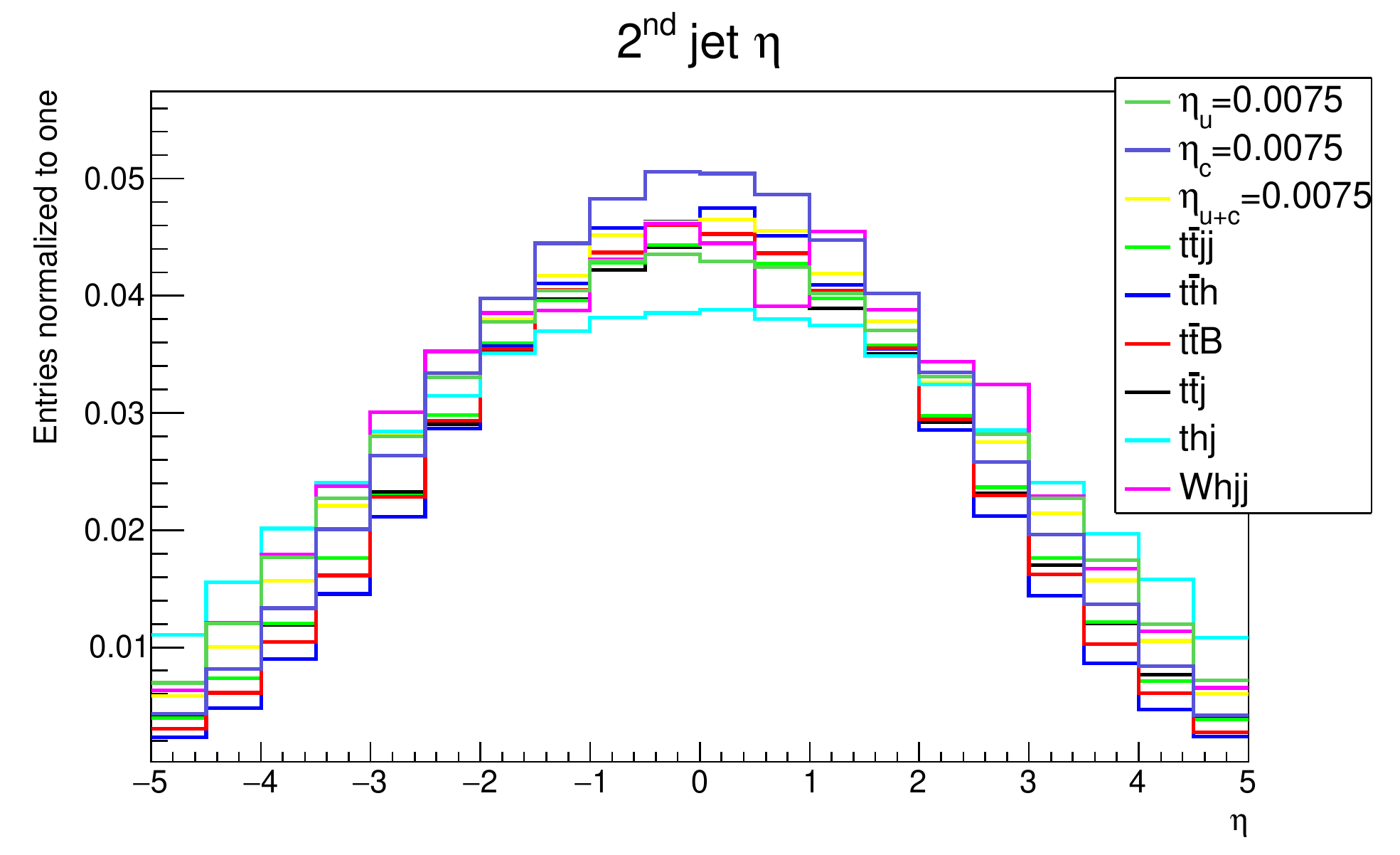}\caption{$2^{\mathrm{nd}}$ jet $\eta$ distribution at detector which are
mostly central.\label{fig:-jet-eta2}}
\end{figure}

\begin{figure}[h]
\includegraphics[scale=0.48]{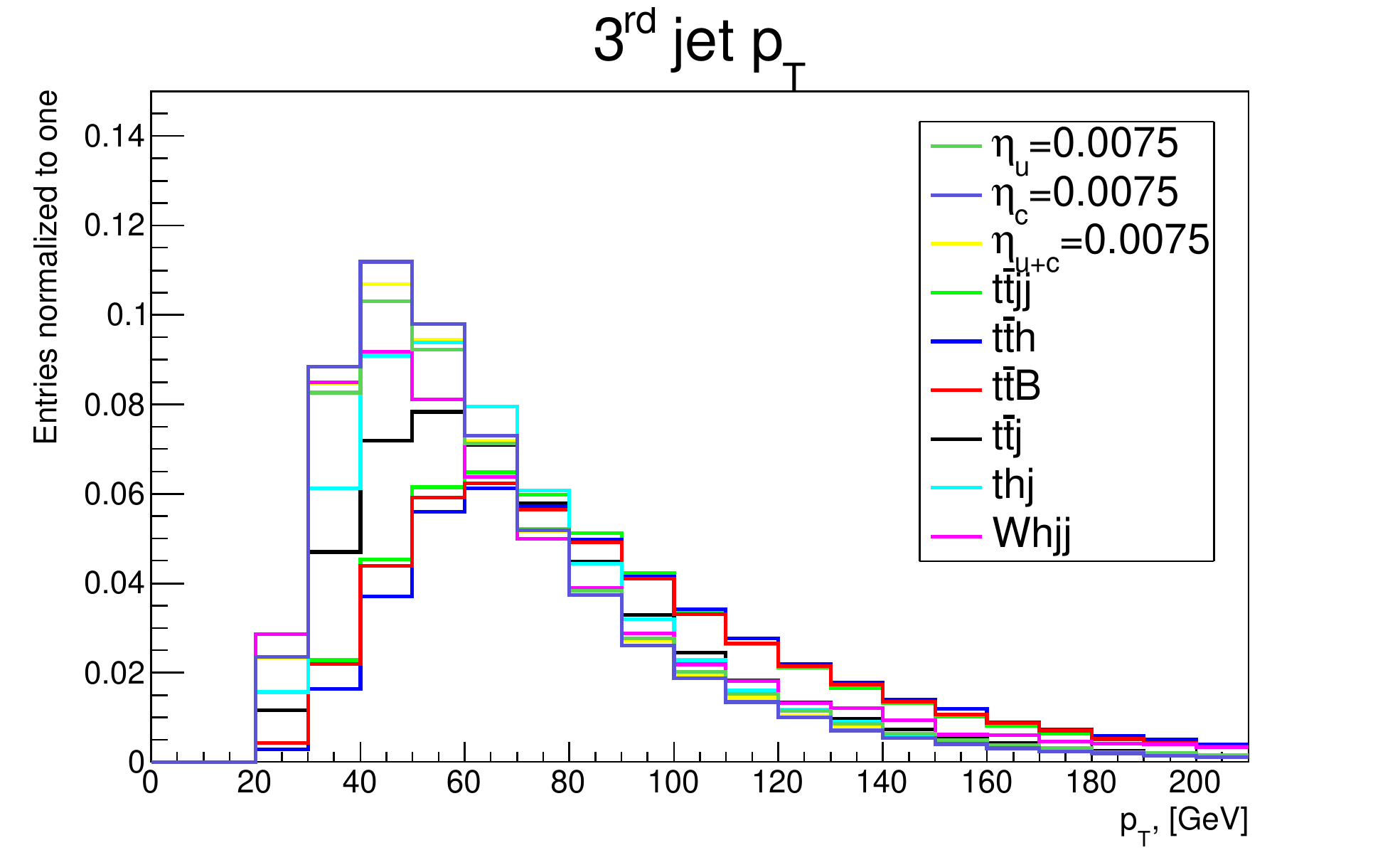}\caption{$3^{\mathrm{rd}}$ jet $p_{T}$ distribution. Similar to $2^{\mathrm{nd}}$jet,
however after tagging, there arise a $p_{T}$ gap between these jets
and the peak is shifted to low $p_{T}$ region.\label{fig:-jet-pt3}}
\end{figure}

\begin{figure}[h]
\includegraphics[scale=0.48]{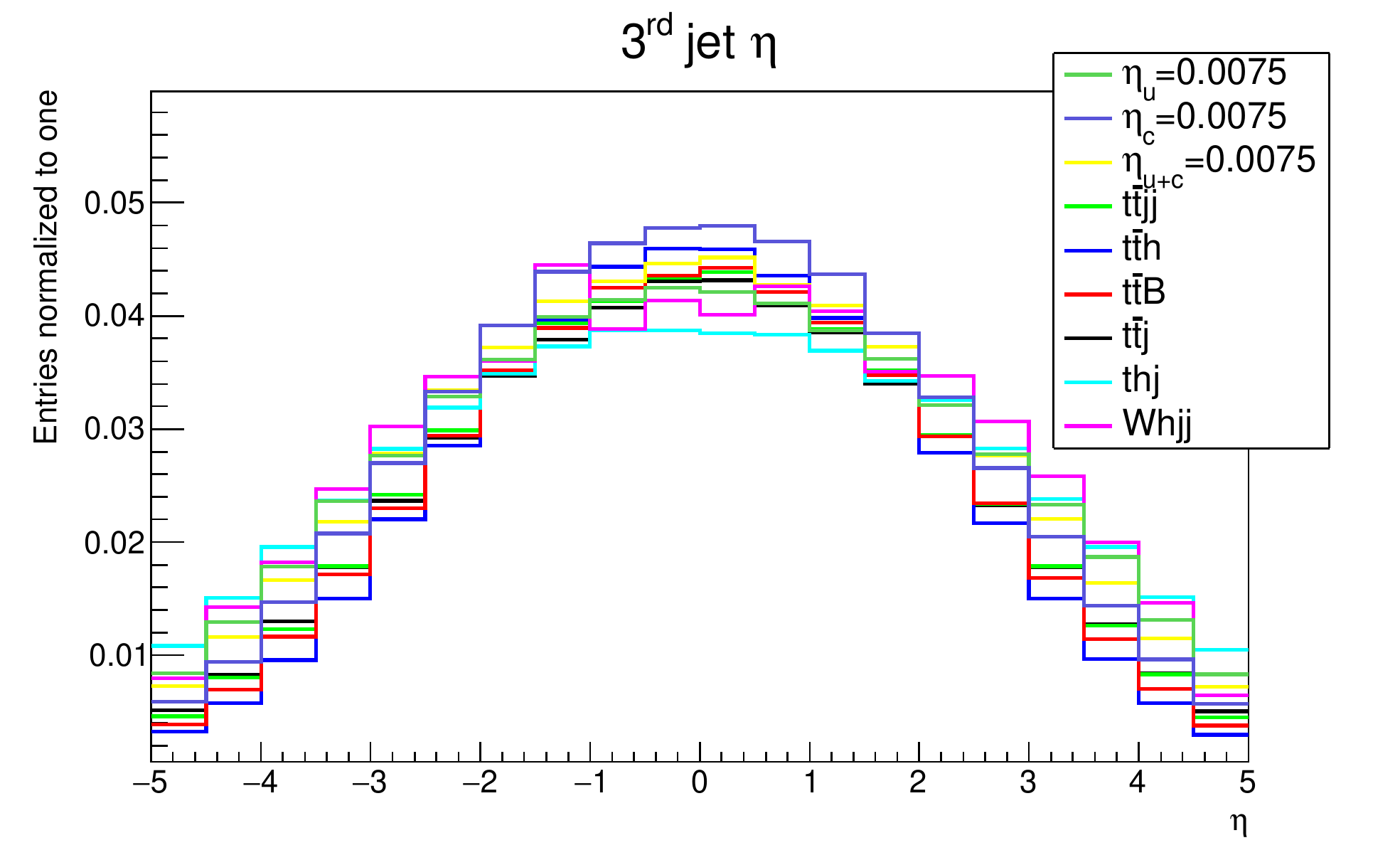}\caption{$3^{\mathrm{rd}}$ jet $\eta$ distribution at detector is central.\label{fig:-jet-eta3}}
\end{figure}

\begin{figure}[h]
\includegraphics[scale=0.48]{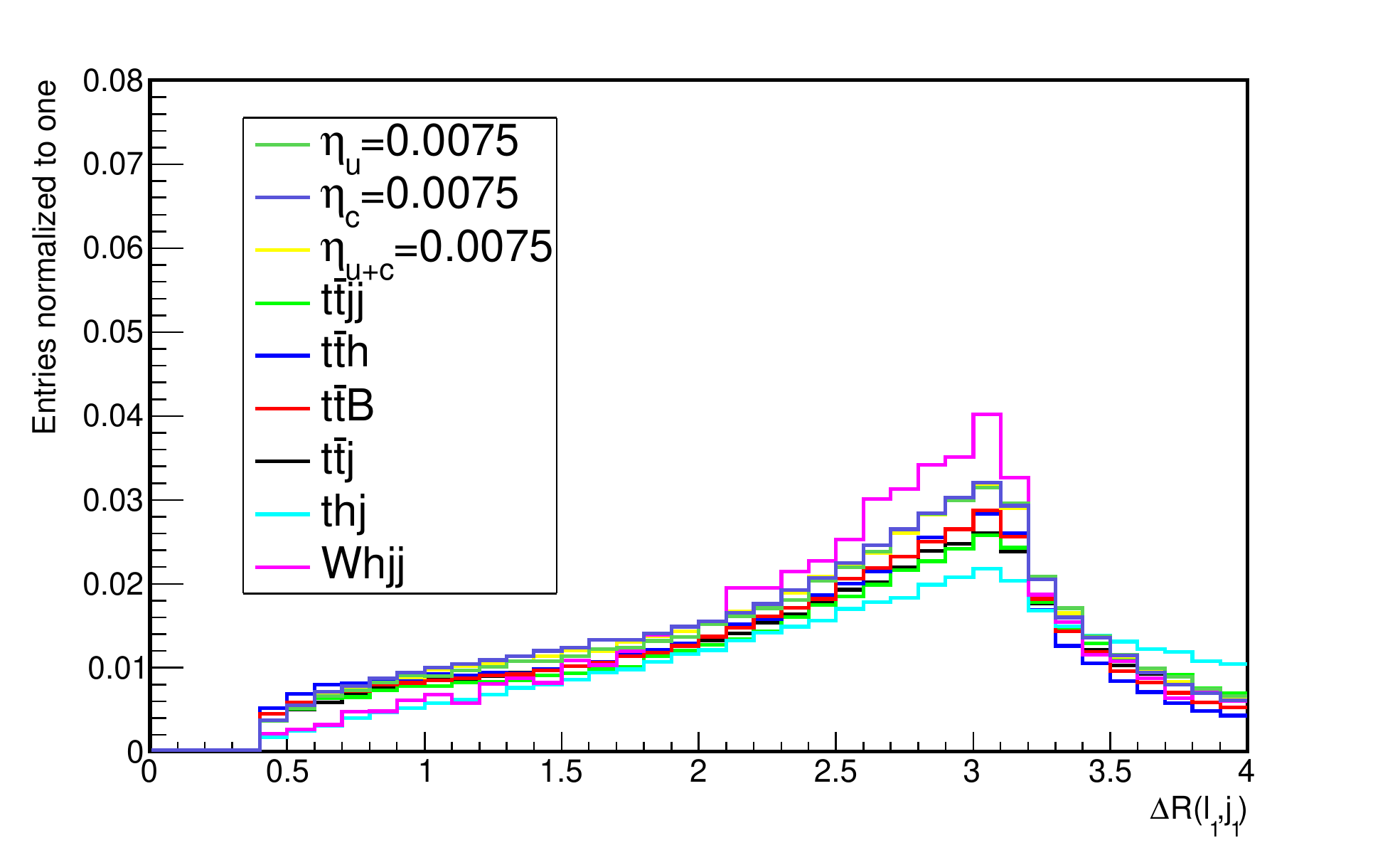}\caption{$\Delta R$ distribution between leading jet and leading lepton\label{fig:-distribution-between leading jet and leading lepton}}
\end{figure}

\begin{figure}[h]
\includegraphics[scale=0.48]{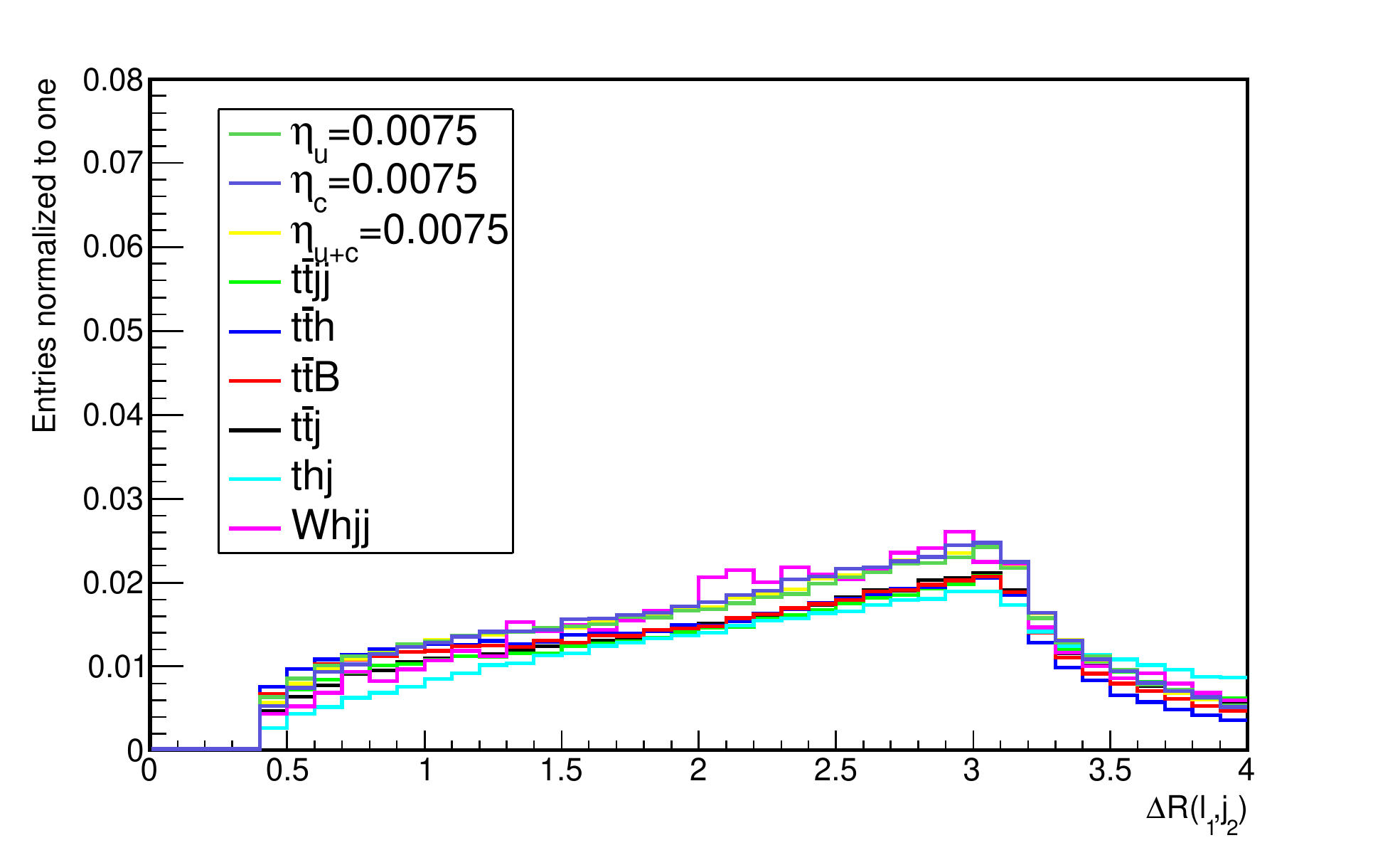}\caption{$\Delta R$ distribution between second leading jet and leading lepton\label{fig:-distribution between second leading jet and leading lepton}}
\end{figure}

\begin{figure}[h]
\includegraphics[scale=0.48]{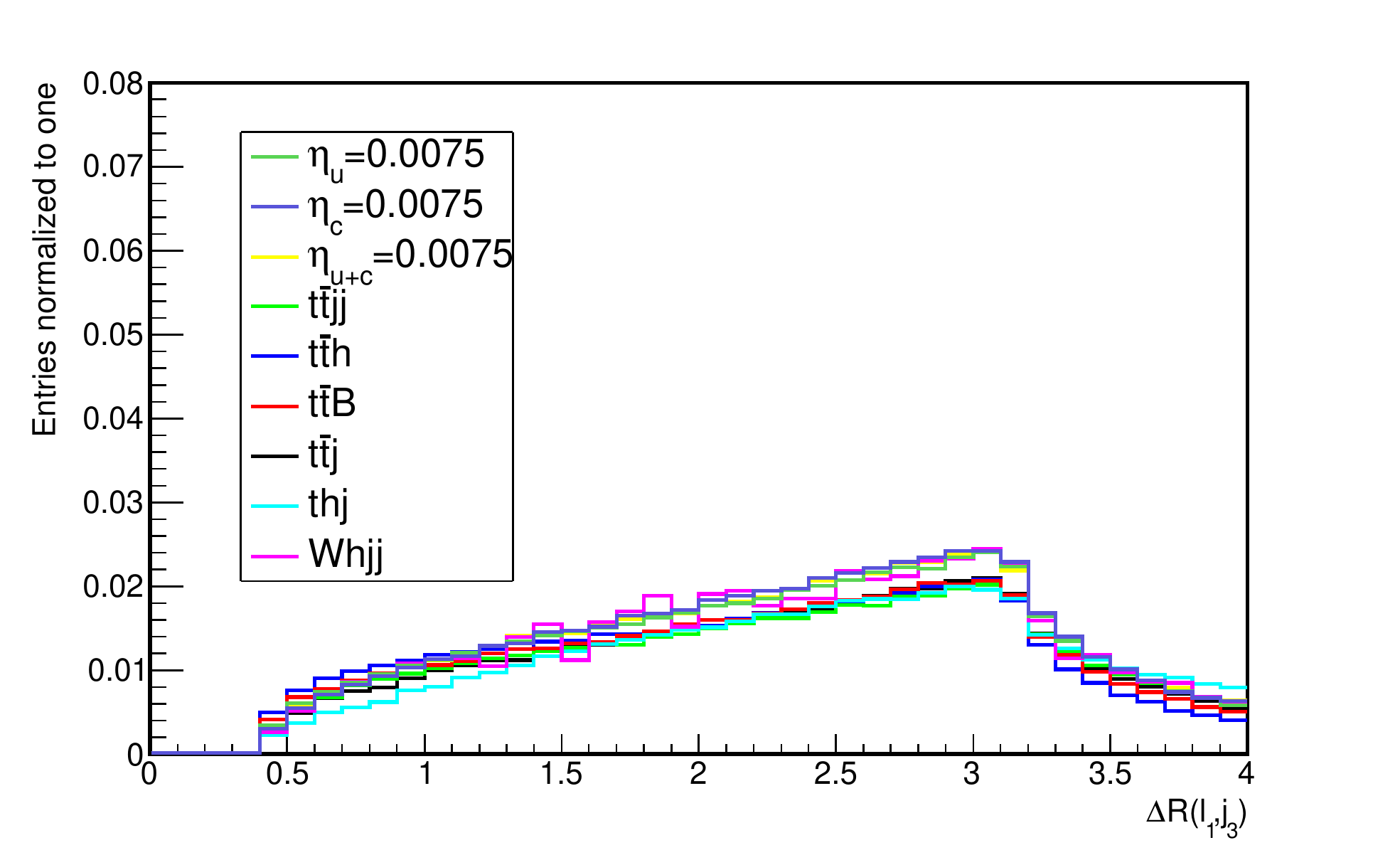}\caption{$\Delta R$ distribution between third leading jet and leading lepton\label{fig:-distribution between third leading jet and leading lepton}}
\end{figure}

\begin{figure}[h]
\includegraphics[scale=0.48]{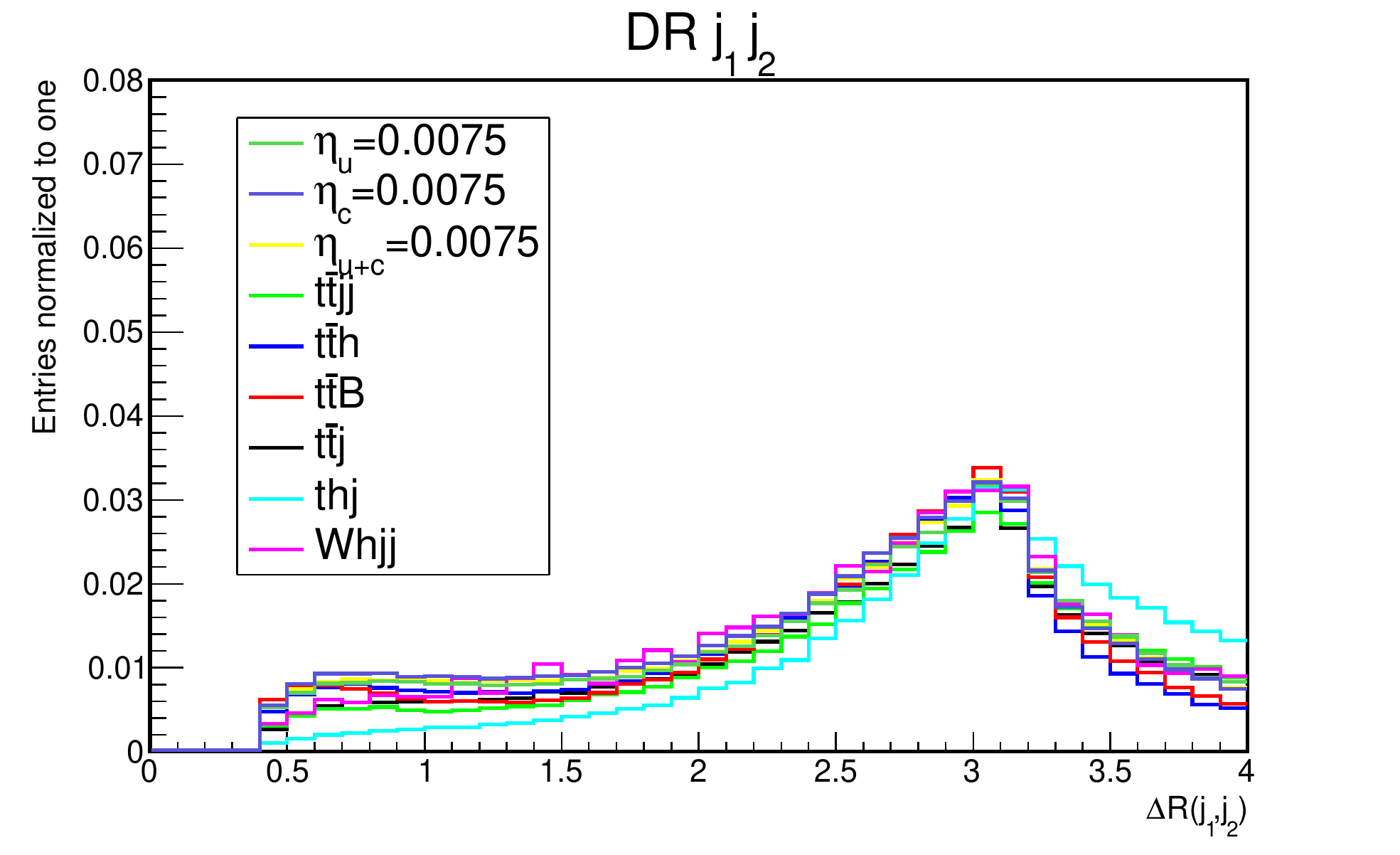}\caption{$\Delta R$ distribution between leading jet and second leading jet\label{fig:-distribution between leading jet and second leading jet}}
\end{figure}

\begin{figure}[h]
\includegraphics[scale=0.48]{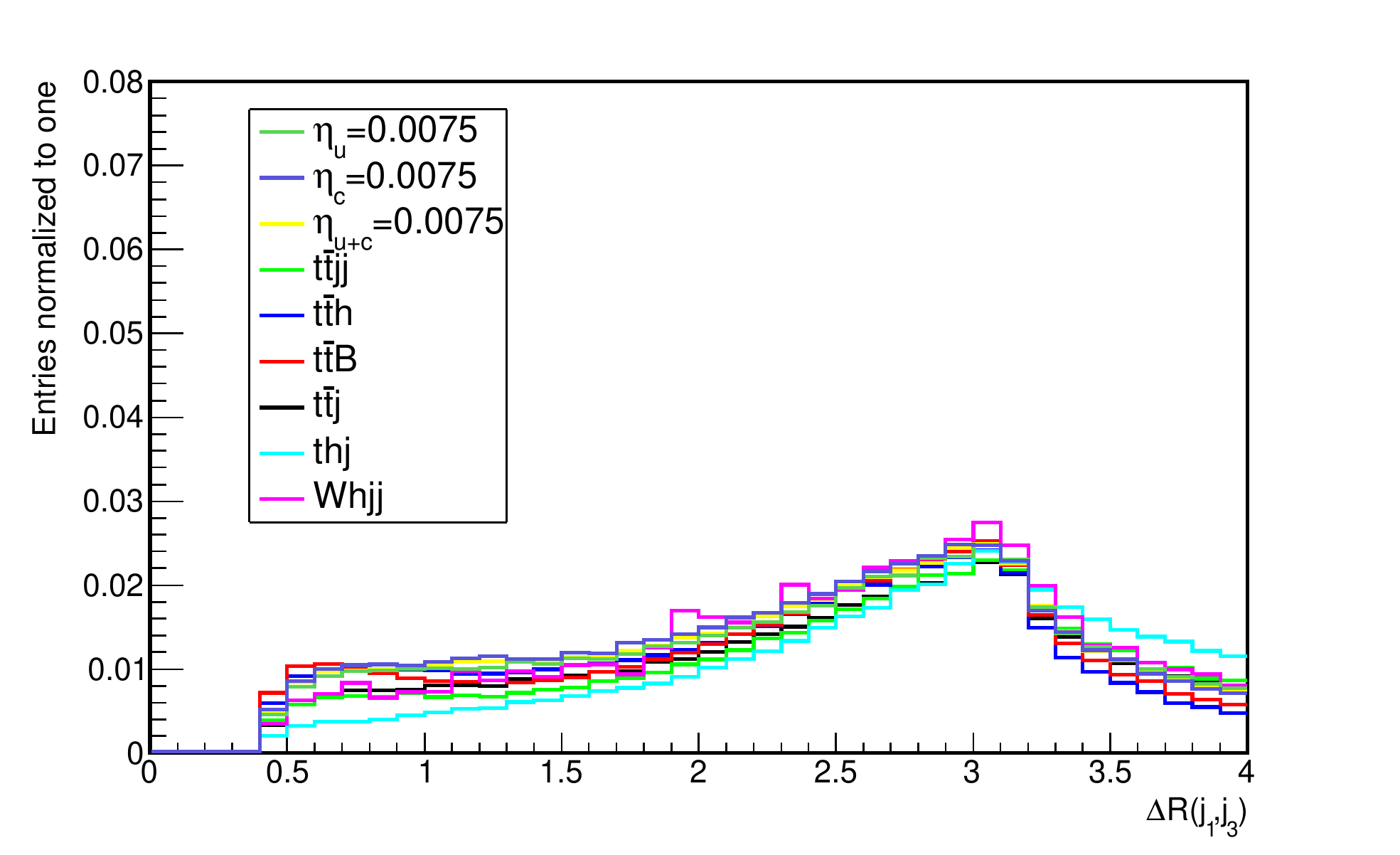}\caption{$\Delta R$ distribution between leading jet and third leading jet\label{fig:-distribution between leading jet and third leading jet}}
\end{figure}

\begin{figure}[h]
\includegraphics[scale=0.48]{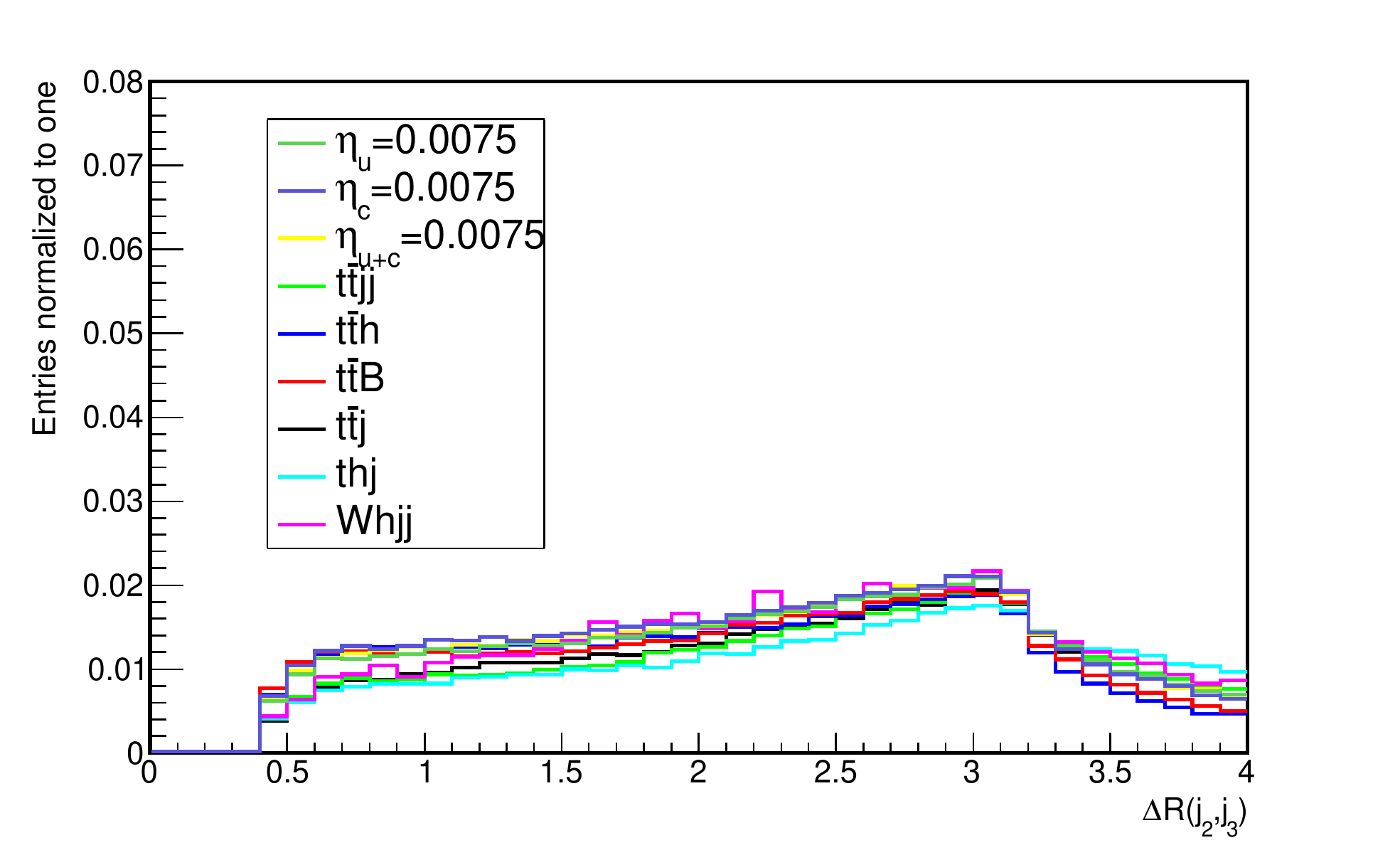}\caption{$\Delta R$ distribution between second leading jet and third leading
jet\label{fig:-distribution between second leading jet and third leading jet}}
\end{figure}

Bearing in mind we have one lepton, MET and three b-jets as final
state in the signal events. In addition to this, remembering the interation
occures at a collider which has 100 TeV COM energy, we need to put
basic cuts firsty to work with good objects rather than irrelevant
random particles. First of all we would like to start with giving
the distributions only generator level cuts present at Fig. \ref{fig:Lepton--distributions},
\ref{fig:MET-distribution-for}, \ref{fig:-jet-:pt1}, \ref{fig:-jet-eta1},
\ref{fig:-jet-:pt2}, \ref{fig:-jet-eta2}, \ref{fig:-jet-pt3}, \ref{fig:-jet-eta3},
\ref{fig:-distribution-between leading jet and leading lepton}, \ref{fig:-distribution between second leading jet and leading lepton},
\ref{fig:-distribution between third leading jet and leading lepton},
\ref{fig:-distribution between leading jet and second leading jet},
\ref{fig:-distribution between leading jet and third leading jet},
\ref{fig:-distribution between second leading jet and third leading jet}.
Then we choose basic cuts as shown Table \ref{tab:List-of-basic}.
We use $\sum p_{T}$ as the sum of $p_{T}$ of the objects which are
used at reconstruction of top and Higgs. 

Defining good objects by using these cuts in a looser manner to keep
number of signal events high at first stage to see behaviour of signal
and background; then we will use more strict cuts to finalize our
analysis while optimise the signal significance. At Table \ref{tab:List-of-basic}
first two lines show our object selection criteria. MET cut has been
included for fullfill event selection criteria as well (then restricted
to work with detected objects instead at main part of analysis). Then
we use the more sensitive regions of detector by applying $\eta$
cuts. $p_{T}$ cuts included for eliminating irrelevant jets and events
basically. Here the $p_{T}$ cuts somewhat self explanatory: we would
like to trim resudual jets produced at interaction and mixes the signal
according to their decay mechanism. Finally $\Delta R$ cuts are present
here for fat jets for possible misleading detection and distorting
our analysis. After we see the bare distributions, we continue with
strengthen our cuts. To show the evolution of the analysis we would
like to divide analysis up to a looser Higgs mass reconstruction criteria
with $\chi^{2}$ method then sharpen our cuts and pass to second part
of analysis to give the last shape to analysis. 

Notice that the background processes $pp\rightarrow t\bar{t}j$ and
$pp\rightarrow t\bar{t}jj$ have highest cross section and has no
Higgs as a product. Even though these backgrounds are inseparable
completely in that sense. We can promote Higgs reconstruction as a
discriminator for analysis. We may look for additional criterion to
impose for rejecting that $t\bar{t}$ pair and furthermore reduce
all backgrounds. This is why we choose Higgs reconstruction as a pin
point of analysis. 

We apply a $\chi^{2}$ which reconstructs Higgs bosons invariant mass
in a range with $\pm20$ GeV arround 125 GeV. To determine the channel
properly, we demand at least one of the jets reconstructs Higgs boson
must be b-tagged. That b-tagging has vital importance to deduce the
channel and a strong restriction to the other jet when we consider
both mass reconstruction and b-tagging. 

Here we used the relation when we calculate the $\chi^{2}$ basically
\begin{equation}
\chi^{2}=\Sigma_{i}\frac{\text{\ensuremath{\left(O_{i}-E\right)}}^{2}}{\Delta^{2}}.
\end{equation}
In this relation we used $O_{i}$ as observed reconstruction variable
for signal and backgrounds at every event. $E$ stands for expected
values of these quantities as usual. $\Delta$ shows the error at
this event reconstruction. If we adopt this relation to our analysis
for the simpler case we may write this relation down as 
\begin{equation}
\chi^{2}=\frac{\left(m_{jj}-m_{H}\right)^{2}}{\sigma_{H}^{2}}
\end{equation}

After we draw a frame for analysis, for a better understanding of
cuts we present important kinematical variables with their entries
normalized to one and comment on them (here only generator level cuts
included).
\begin{figure}[h]
\raggedright{}\includegraphics[scale=0.48]{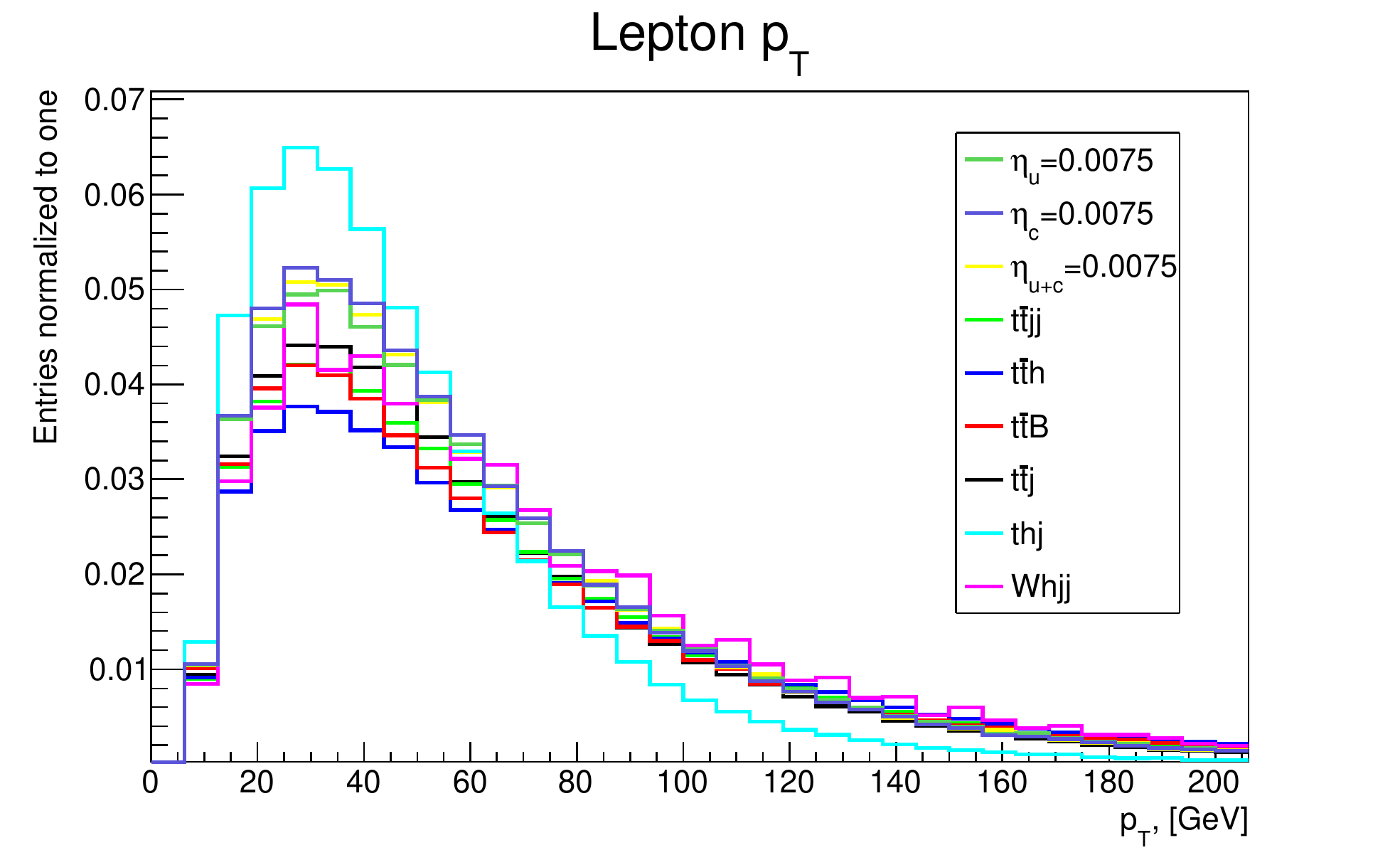}\caption{Because of high $p_{T}$ of top quark and relavant decay chains, lepton
behaviour is boosted. Besides they have a peak arround 30 GeV implies
that their production mechanism is disintegration of $W$ bosons.\label{fig:Because-of-high}}
\end{figure}
We present histograms showing the characteristics of jets produced
and little comment on them, see Fig. \ref{fig:-jet-:pt1}, \ref{fig:-jet-eta1},
\ref{fig:-jet-:pt2}, \ref{fig:-jet-eta2}, \ref{fig:-jet-pt3}, \ref{fig:-jet-eta3}.
\begin{table}[h]
\raggedright{}\caption{\label{tab:List-of-basic}Region selection and basic cuts for pre-analysis
and analysis: Difference occurs at analysis part when an additional
MET<100 GeV cut applied. }
\begin{tabular}{|c|}
\hline 
Region selection and basic cuts\tabularnewline
\hline 
\hline 
$N(jets)\geq3$\tabularnewline
\hline 
$N(l^{\pm})=1$\tabularnewline
\hline 
$p_{T}^{j}>30\ \mathrm{GeV}$\tabularnewline
\hline 
$p_{T}^{l}>25\ \mathrm{GeV}$\tabularnewline
\hline 
$MET>30\ \mathrm{GeV}$\tabularnewline
\hline 
$|\eta^{l,j}|<3$\tabularnewline
\hline 
$\Delta R(i,j)_{\mathrm{All}}>0.4$\tabularnewline
\hline 
\end{tabular}
\end{table}

\begin{figure}[h]
\includegraphics[scale=0.48]{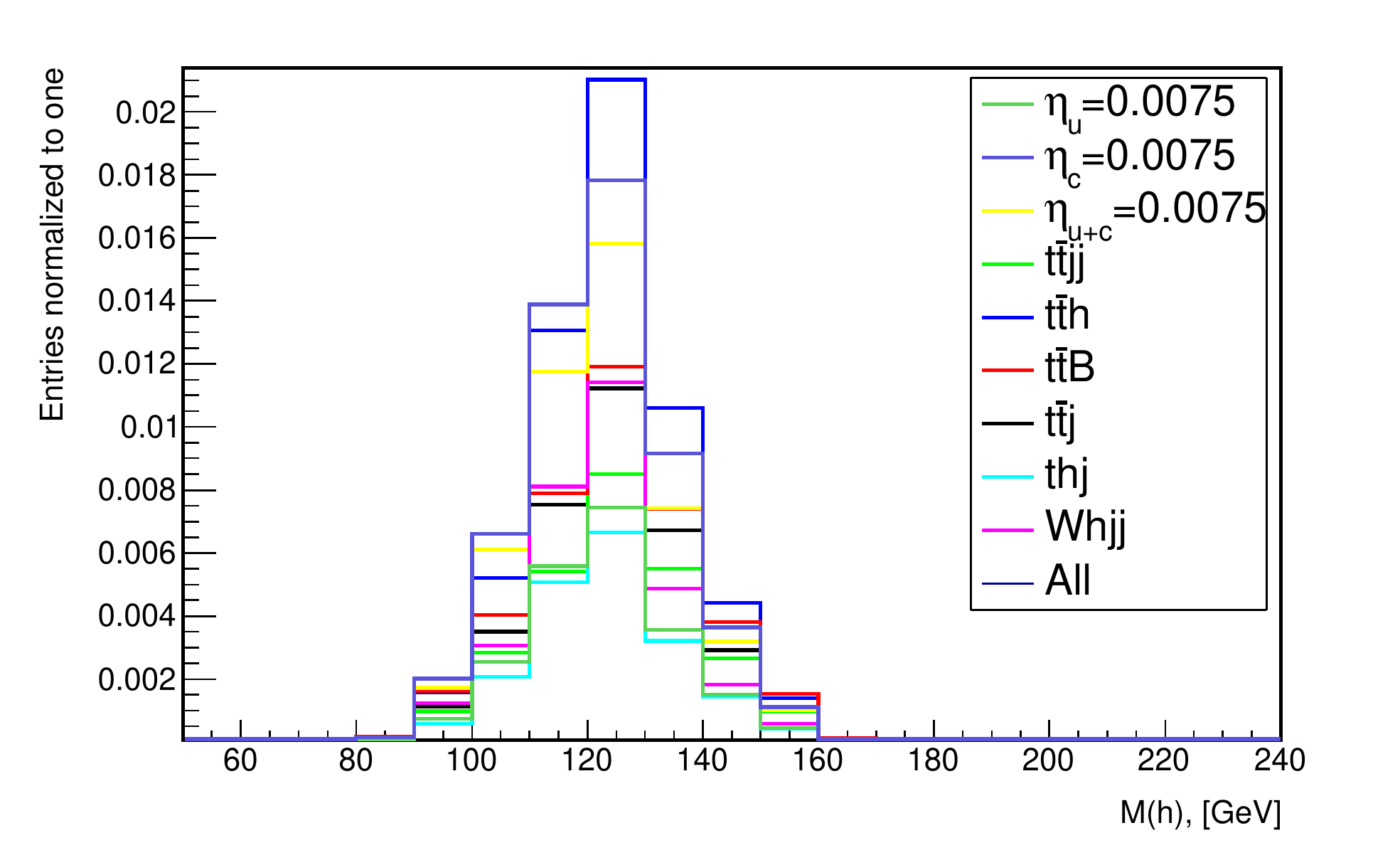}

\caption{Higgs invariant mass distribution after pre-analysis. \label{fig:Higgs-invariant-mass}}

\end{figure}

\begin{figure}[h]
\includegraphics[scale=0.48]{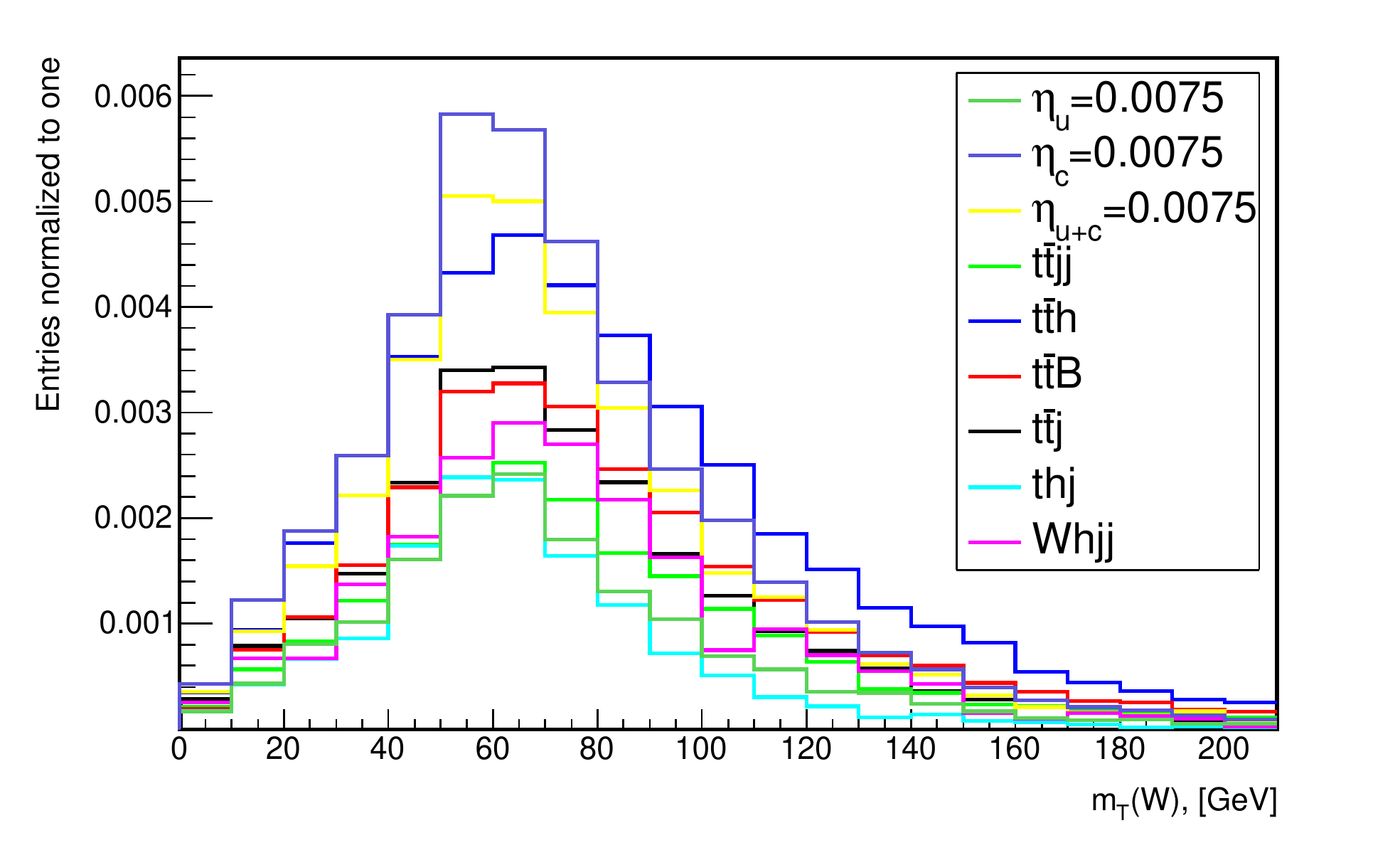}\caption{$W$ boson transverse mass distribution after pre-analysis.\label{fig:-boson-transverse}}

\end{figure}

\begin{figure}[h]
\includegraphics[scale=0.48]{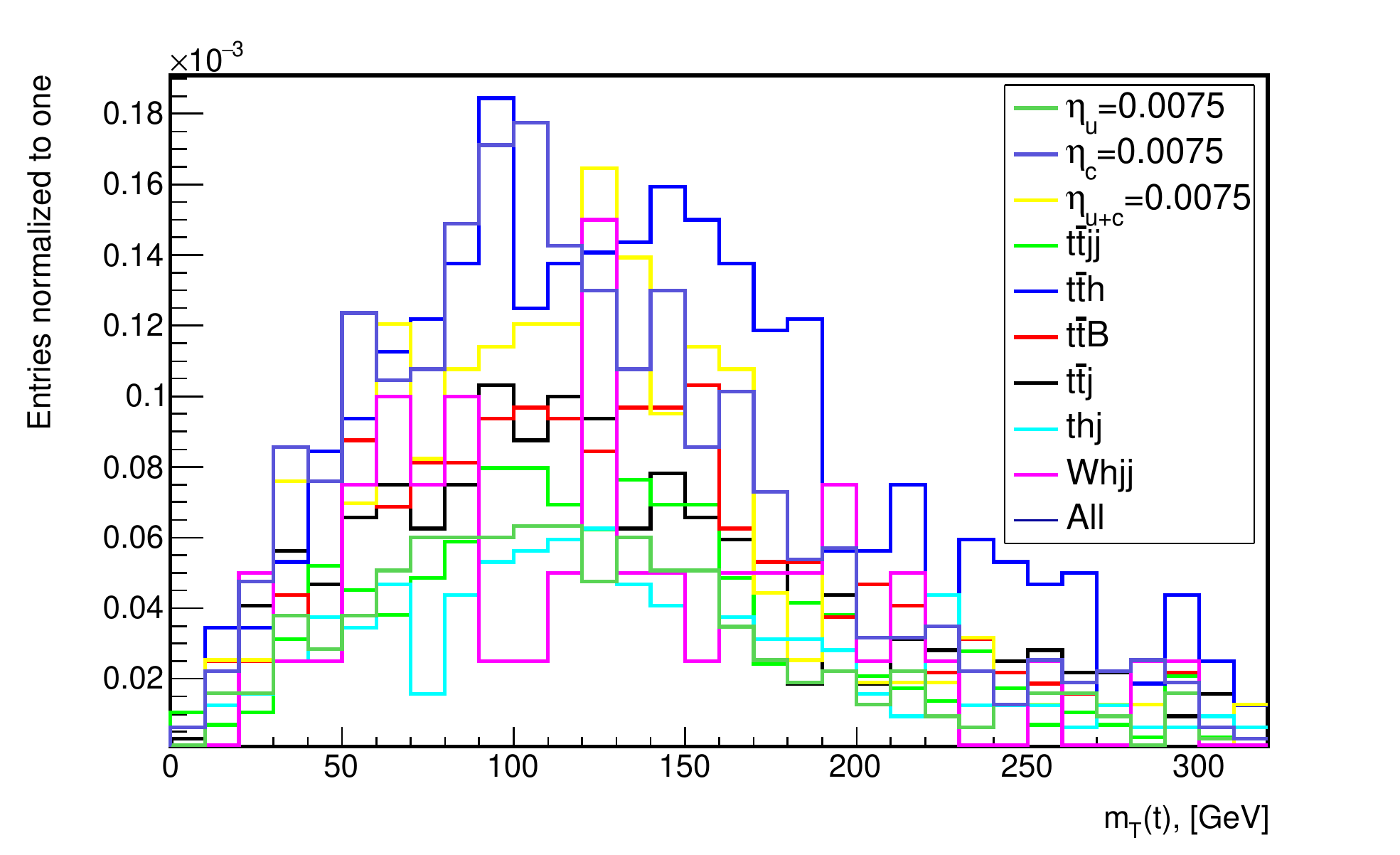}

\caption{Top quark transverse mass distribution after pre-analysis.\label{fig:Top-quark-transverse}}

\end{figure}

\begin{figure}[h]
\includegraphics[scale=0.48]{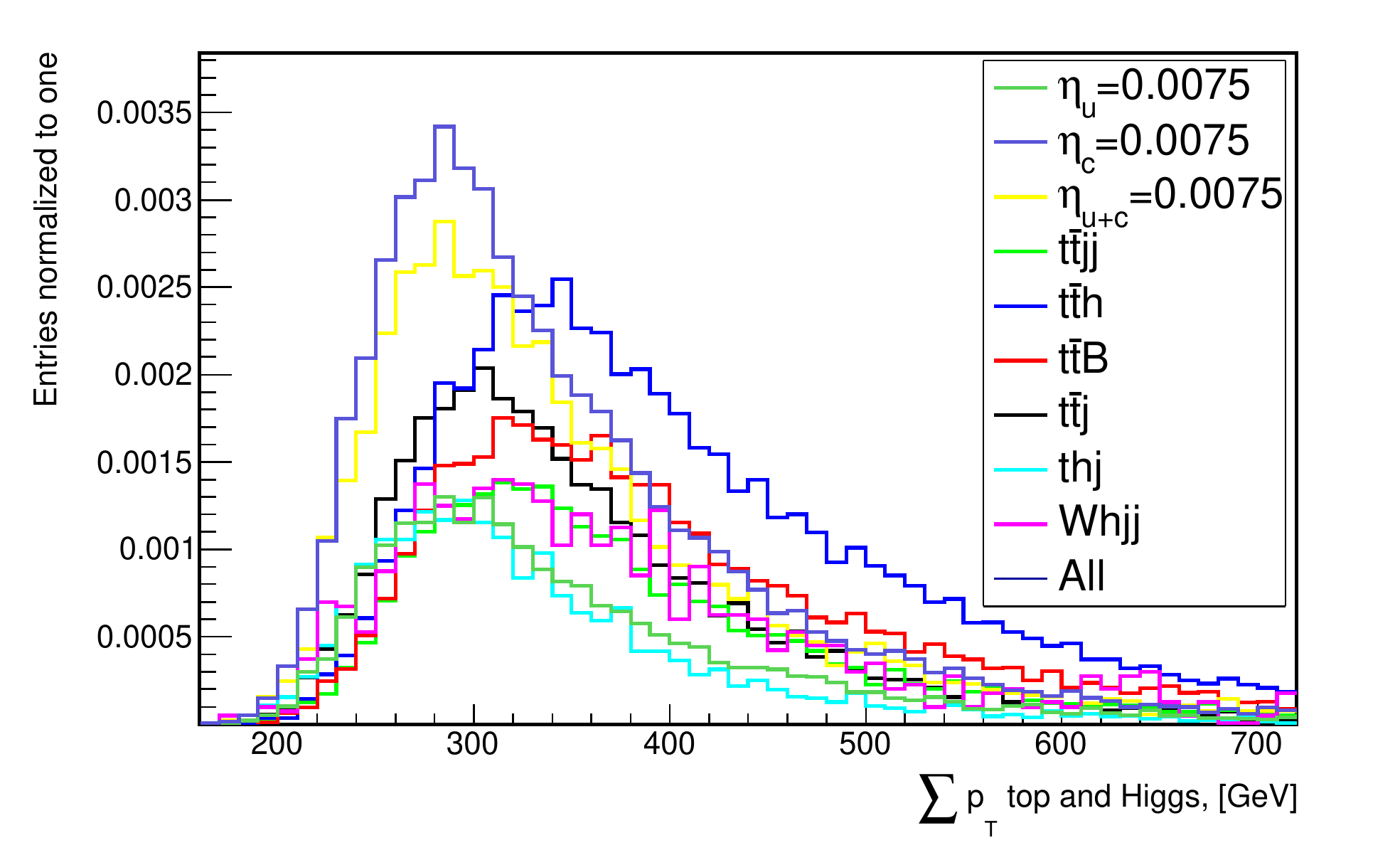}

\caption{$p_{T}$ sum of final state object which are used to reconstruct Higgs
and top quark after pre-analysis. \label{fig:-sum-of}}

\end{figure}

That simple approach depends on the best reconstruction of Higgs and
relatively the other final state objects with a large event number
and in a wide distribution range, fullfilling the true decay channel.
This pre-analysis proofs our expectations about the appling cuts in
a more sophiticated way and also inseperability of signal and background
even with a fundamental level of $\chi^{2}$ and relavant physical
conditions. As you will see from Fig. \ref{fig:Higgs-invariant-mass},
\ref{fig:-boson-transverse}, \ref{fig:Top-quark-transverse}, \ref{fig:-sum-of}
the gap between at number of events of signal and background is fairly
high. One gets no relavant signal significance for such an analysis.
It is reflacted at kinematic and derived variables, such that there
is no region to seperate signal and background by applying cuts to
these which is transfered to final histograms. We would like to give
a cut-efficiency table for this pre-analysis to state our analysis
path clearly.
\begin{table}[h]

\caption{Cut efficiencies for signal and background processes for pre-analysis.}

\raggedright{}%
\begin{tabular}{|c|c|c|c|}
\hline 
\multirow{2}{*}{Process} & Region & \multirow{2}{*}{Basic cuts (\%)} & \multirow{2}{*}{$\chi^{2}$ (\%)}\tabularnewline
 & selection (\%) &  & \tabularnewline
\hline 
$pp\rightarrow t(\bar{t})h$ & \multirow{3}{*}{$64.07$} & \multirow{3}{*}{$15.98$} & \tabularnewline
$pp\rightarrow t(\bar{t})hj$ &  &  & $4.72$\tabularnewline
$\eta_{u}=\eta_{c}=0.0075$ &  &  & \tabularnewline
\hline 
$pp\rightarrow t\bar{t}j$ & $56.57$ & $15.91$ & $3.43$\tabularnewline
\hline 
$pp\rightarrow t\bar{t}jj$ & $56.79$ & $17.06$ & $2.70$\tabularnewline
\hline 
$pp\rightarrow t\bar{t}h$ & $55.34$ & $21.33$ & $5.75$\tabularnewline
\hline 
$pp\rightarrow t\bar{t}B$ & $55.93$ & $19.69$ & $3.85$\tabularnewline
\hline 
$pp\rightarrow thj$ & $63.59$ & $9.16$ & $1.95$\tabularnewline
\hline 
$pp\rightarrow Whjj$ & $64.59$ & $14.80$ & $3.12$\tabularnewline
\hline 
\end{tabular}
\end{table}

In the folowing paragraphs we present our analysis path and introduce
signal significance relation. We will study here for three cases and
find search limits seperately for $\eta_{u+c}$, $\eta_{u}$ and $\eta_{c}$
as promised.

The main part of our analysis is to reconstruct Higgs mass and top
quark transverse mass (while considering $W$ bosons transverse mass)
from final state objects mentioned before in the text. For a complete
analysis we must recover both of them, since they are charactaristics
of our signal process. Nevertheless, a top quark with a leptonic decay
can be reconsturcted by its $m_{T}$. There is a missing knowledge
due to longitudinal component of neutrino momentum. On the other side
we can reconstruct Higgs boson using its invariant mass more precisely.
Then we offer a chi square $(\chi^{2})$ that must restrict Higgs'
invariant mass considerably in comparison to tops $m_{T}$ which can
stay looser. 

Another important aspect of reconstruction of a top quark and a Higgs
is to use b-tagging mentioned as long before. Since we are considering
$h\rightarrow b\bar{b}$, hencefort at least one or if possible two
jets that reconstructs Higgs must be b-tagged. Seperately, when we
come to reconstruction of top quark one b-jet is is needed at first
glance. However, remembering only source of lepton and MET is top
decay for signal case and we can also restrict top quarks transverse
mass additionally. Both event selection criteria, $\chi^{2}$ and
$m_{T}$ limitations cancels this two or three b-tag necessity ($W$
boson's tansverse mass is also restricted) and optimise analyse greatly.
We just simply impose these cuts at our anlysis code. We use 
\begin{align}
m_{T}^{t} & =\left(\sqrt{\left(p^{l}+p^{b}\right)^{2}+\left|\vec{p}_{T}^{l}+\vec{p}_{T}^{b}\right|^{2}}+\left|\vec{p}_{T}^{\nu_{l}}\right|\right)^{2}\nonumber \\
 & -\left|\vec{p}_{T}^{l}+\vec{p}_{T}^{b}+\vec{p}_{T}^{\nu_{l}}\right|^{2}
\end{align}
for reconstructing top quarks transverse mass and for transverse mass
for $W$ boson
\begin{equation}
m_{T}^{W}=\sqrt{2p_{T}^{l}E_{T}^{\mathrm{miss}}-\vec{p}_{T}^{l}.\vec{p}_{T}^{\nu_{l}}}.
\end{equation}

\begin{itemize}
\item Cut 1: $30<m_{T}^{W}<90$, transverse mass restriction of $W$ for
better event reconstruction from final state objects. 
\item Cut 2: $100<m_{T}^{t}<200$ transverse mass restriction of top quark
for better event reconstruction from final state objects. 
\end{itemize}
We observe a clear decrease in our main backgrounds, but analysis
is still incomplete and results must be eleborated to compare with
other studies in literature. Main problem is $t\bar{t}+\mathrm{jet(s)}$,
must be fully eliminated at some region where signal event has dominance
that gives meaningful signal significance value. Besides the other
backgrounds still problematic, since elimation is rather difficult
due to presence of bosons, besides $pp\rightarrow thj$ process is
directly present at SM too. So we need to discriminate them. 

To complete analysis we use much more sophisticated variables then
use $\Delta R$ cuts considering physics between the objects we are
interested in.
\begin{itemize}
\item Cut 3: Contransverse mass \citep{key-99} using the jets that reconstruct
Higgs boson as defined as
\begin{equation}
m_{CT}=\sqrt{p_{T}^{b_{1}}p_{T}^{b_{2}}\cos(1+\Delta\phi_{b_{1}b_{2}})}
\end{equation}
and we demand $m_{CT}<400$ GeV. This cut targets mainly $t\bar{t}+$jet(s)
backgrounds but also effective on the others. By applying this cut
we are aiming to put $t\bar{t}+$jet(s) backgrounds in pieces and
open a way to use $\Delta R$ cuts to trim out these.
\item Cut 4: As we know Higgs and additional jet comes from a virtual top
quark at signal process we may deal with a hadronic reconstruction
of this particle. Although it is virtual particle and its distribution
cannot give a sharp peak arround its invariant mass, we still get
a splayed distribution roughly arround 172 GeV. When the subject is
reconstruction of signal event from reverse, to narrow the reconstruction
circle we apply additional $\chi^{2}$ as such
\begin{equation}
\chi^{2}=\frac{\left(m_{j_{1}j_{2}j_{3}}-m_{\mathrm{top}}\right)^{2}}{\sigma_{top}^{2}}
\end{equation}
taking $\sigma=20$ GeV. By applying this cut we finally extract different
$\Delta R$ behaviours of siganland backgrounds, so we can start to
trim out regions to maximize signal significance.
\item Cut 5: When we consider $pp\rightarrow th(j)$ process, the most favorable
momentum configuration is the back to back scatter of top quark and
Higgs which resulted with the cut $2.2<\Delta R(t,h)<3$. We use the
objects which are recognized as top cluster and Higgs cluster after
tagging jets accordingly by using previous cuts. 
\item Cut 6: Same $\Delta R$ corelation holds between ``jet from top''
and Higgs cluster too. Thus we seek to observe $\Delta R(j_{t},h)>2$.
\item Cut 7: After leptonic decay of top quark, lepton and jet is back to
back at their rest frame. However, as boosted objects they must be
seperated enough as such, $\Delta R(j_{t},l)>1.8$.
\item Cut 8: The corelation at the $\Delta R$ variable between top and
Higgs cluster partly transfered to final state objects especially
the lepton and a ``jet from Higgs'', so we demand $\Delta R(j_{h},l)>2$.
\end{itemize}
As one can expect the numbers at $\Delta R$ cuts selected to optimise
signal significance. Here we would like to give cut efficiency at
Tab. \ref{tab:Cut-efficiencies-for}. 
\begin{table}[h]
\raggedright{}\caption{Cut efficiencies for signal and background processes given \% ratios.
Signal choosen as scenario $\eta_{u}=\eta_{c}=0.0075$. \label{tab:Cut-efficiencies-for}}
\begin{tabular}{|c|c|c|c|c|c|c|c|}
\hline 
 & {\footnotesize{}Signal} & {\footnotesize{}$t\bar{t}j$} & {\footnotesize{}$t\bar{t}jj$} & {\footnotesize{}$t\bar{t}h$} & {\footnotesize{}$t\bar{t}B$} & {\footnotesize{}$thj$} & {\footnotesize{}$Whjj$}\tabularnewline
\hline 
{\footnotesize{}Region} & \multirow{2}{*}{{\footnotesize{}64.07}} & \multirow{2}{*}{{\footnotesize{}56.57}} & \multirow{2}{*}{{\footnotesize{}56.79}} & \multirow{2}{*}{{\footnotesize{}55.34}} & \multirow{2}{*}{{\footnotesize{}55.92}} & \multirow{2}{*}{{\footnotesize{}63.59}} & \multirow{2}{*}{{\footnotesize{}64.59}}\tabularnewline
{\footnotesize{}selection} &  &  &  &  &  &  & \tabularnewline
\hline 
{\footnotesize{}Basic} & \multirow{2}{*}{{\footnotesize{}12.17}} & \multirow{2}{*}{{\footnotesize{}11.42}} & \multirow{2}{*}{{\footnotesize{}11.37}} & \multirow{2}{*}{{\footnotesize{}13.50}} & \multirow{2}{*}{{\footnotesize{}12.32}} & \multirow{2}{*}{{\footnotesize{}7.38}} & \multirow{2}{*}{{\footnotesize{}9.88}}\tabularnewline
{\footnotesize{}cuts} &  &  &  &  &  &  & \tabularnewline
\hline 
{\footnotesize{}$\chi_{H}^{2}$} & {\footnotesize{}1.56} & {\footnotesize{}1.10} & {\footnotesize{}0.80} & {\footnotesize{}1.80} & {\footnotesize{}1.00} & {\footnotesize{}0.72} & {\footnotesize{}1.04}\tabularnewline
\hline 
{\footnotesize{}Cut 1} & {\footnotesize{}0.86} & {\footnotesize{}0.58} & {\footnotesize{}0.41} & {\footnotesize{}0.87} & {\footnotesize{}0.50} & {\footnotesize{}0.41} & {\footnotesize{}0.54}\tabularnewline
\hline 
{\footnotesize{}Cut 2} & {\footnotesize{}0.023} & {\footnotesize{}0.011} & {\footnotesize{}0.012} & {\footnotesize{}0.015} & {\footnotesize{}0.0081} & {\footnotesize{}0.069} & {\footnotesize{}0.025}\tabularnewline
\hline 
{\footnotesize{}Cut 3} & {\footnotesize{}0.0089} & {\footnotesize{}0.0038} & {\footnotesize{}0.0055} & {\footnotesize{}0.0062} & {\footnotesize{}0.0031} & {\footnotesize{}0.0038} & {\footnotesize{}0.005}\tabularnewline
\hline 
{\footnotesize{}Cut 4} & {\footnotesize{}0.0070} & {\footnotesize{}0.0025} & {\footnotesize{}0.0031} & {\footnotesize{}0.0025} & {\footnotesize{}0.0016} & {\footnotesize{}0.0022} & {\footnotesize{}0.0025}\tabularnewline
\hline 
{\footnotesize{}Cut 5} & {\footnotesize{}0.0051} & {\footnotesize{}0.0012} & {\footnotesize{}0.0010} & {\footnotesize{}0.00094} & {\footnotesize{}0.00094} & {\footnotesize{}0.0012} & {\footnotesize{}0.0025}\tabularnewline
\hline 
{\footnotesize{}Cut 6} & {\footnotesize{}0.0044} & {\footnotesize{}0.0012} & {\footnotesize{}0.00035} & {\footnotesize{}0.00094} & {\footnotesize{}0.00094} & {\footnotesize{}0.00094} & {\footnotesize{}-}\tabularnewline
\hline 
{\footnotesize{}Cut 7} & {\footnotesize{}0.0019} & {\footnotesize{}-} & {\footnotesize{}-} & {\footnotesize{}0.00062} & {\footnotesize{}0.00062} & {\footnotesize{}0.00094} & {\footnotesize{}-}\tabularnewline
\hline 
{\footnotesize{}Cut 8} & {\footnotesize{}0.0019} & {\footnotesize{}-} & {\footnotesize{}-} & {\footnotesize{}0.00031} & {\footnotesize{}0.00031} & {\footnotesize{}0.00062} & {\footnotesize{}-}\tabularnewline
\hline 
\end{tabular}
\end{table}

Let us comment on cuts and finalize analysis discussion: We started
with selecting the proper final objects and region selection then
use basic cuts to work with relevant objects before reconstruction
of signal process. Selecting one lepton is vital to tag top quark
with $m_{T}^{t}$ restriction to avoid b-tagging. $\chi^{2}$ for
Higgs and tagging one of its constituent jet as $b$ is another important
step for optimising b-tagging while getting full image of both ingredients
of signal process. However, all backgrounds mimic these features,
$t\bar{t}+$jet(s) backgrounds have one leptonic decayed top and many
candidate jets to reconctruct Higss invariant mass being also transparent
to b-tag condition. Their large cross section also enhance many combinations
to increase mixings. Other backgrounds which have bosons too also
additional options coming from bosons and can be pass these filters.
So up to now b-tagging issue and good objects selection is done.

After that point we apply contransverse mass cut and impose $\chi^{2}$
for hadronically decayed virtual top quark. These cuts sharpens Higgs
presence and start to differ signal and background events. Observe
that many backgrounds lack of Higgs. Up to this point all backgrounds
are in game even though their $\Delta R$ distributions started to
become evident. Then we trim out all these differences using the other
cuts and count the number of events at a region where signal events
are concentrated. 

To sum up, we started to apply the cuts where mainly signal oriented,
but also backgrounds have common. We keep track of changes and focus
on Higgs reconstruction and suppress backgrounds to get a viable results.

\section{Results and Conclus\i ons }

Second part of our analysis allows us to eliminate some of backgrounds
at last section. Moreover, the cuts have also suppressed backgrounds
which have large number of events comparing to signal at least several
order of magnitude at generetor level. Now we are ready to evaluate
our results by regarding statistical significance ($SSwS$ stands
for signal significance with systematics) \citep{key-5-1,key-5-2,key-5-3,key-5-4}.
We use the equation
\begin{align}
SSwS_{\mathrm{disc}} & =\left[2\left((S+B)\ln\left(\frac{(S+B)(B+S^{2})}{B^{2}+(S+B)S^{2}}\right)\right.\right.\nonumber \\
 & \left.\left.-\frac{B^{2}}{\Delta_{B}^{2}}\ln\left(1+\frac{\Delta_{B}^{2}S}{B(B+\Delta_{B}^{2})}\right)\right)\right]^{\frac{1}{2}}\label{eq: 4}
\end{align}
for statistical significance in case of inclusion of the systematics.
This reduces to the equation 
\begin{equation}
SS_{\mathrm{disc}}=\sqrt{2[(S+B)\ln(1+S/B)-S]}\label{eq:3}
\end{equation}
when systematic uncertainties neglected for discovery scenario. On
top of that sometimes it is better to exploit the region that a collider
can exlude, for this we use the equation
\begin{align}
SSwS_{\mathrm{exc}}= & \left[2\left\{ S-B\ln\left(\frac{B+S+x}{2B}\right)\right.\right.\nonumber \\
 & \left.-\frac{B^{2}}{\Delta_{B}^{2}}\ln\left(\frac{B-S-x}{2B}\right)\right\} \nonumber \\
 & \left.-\left(B+S-x\right)\left(1+\frac{B}{\Delta_{B}^{2}}\right)\right]^{1/2}\label{eq:5}
\end{align}
where 

\begin{equation}
x=\sqrt{\left(S+B\right)^{2}-\frac{4SB\Delta_{B}^{2}}{B+\Delta_{B}^{2}}}
\end{equation}
and $\Delta_{B}$ corresponds to uncertainty of background events;
at limiting case $\Delta_{B}=0$ Eq. \ref{eq:5} reduces to following
equation

\begin{equation}
SS_{\mathrm{exc}}=\sqrt{2\left(S-B\ln\left(1+\frac{S}{B}\right)\right).}\label{eq:6}
\end{equation}

\begin{figure}[h]
\includegraphics[scale=0.44]{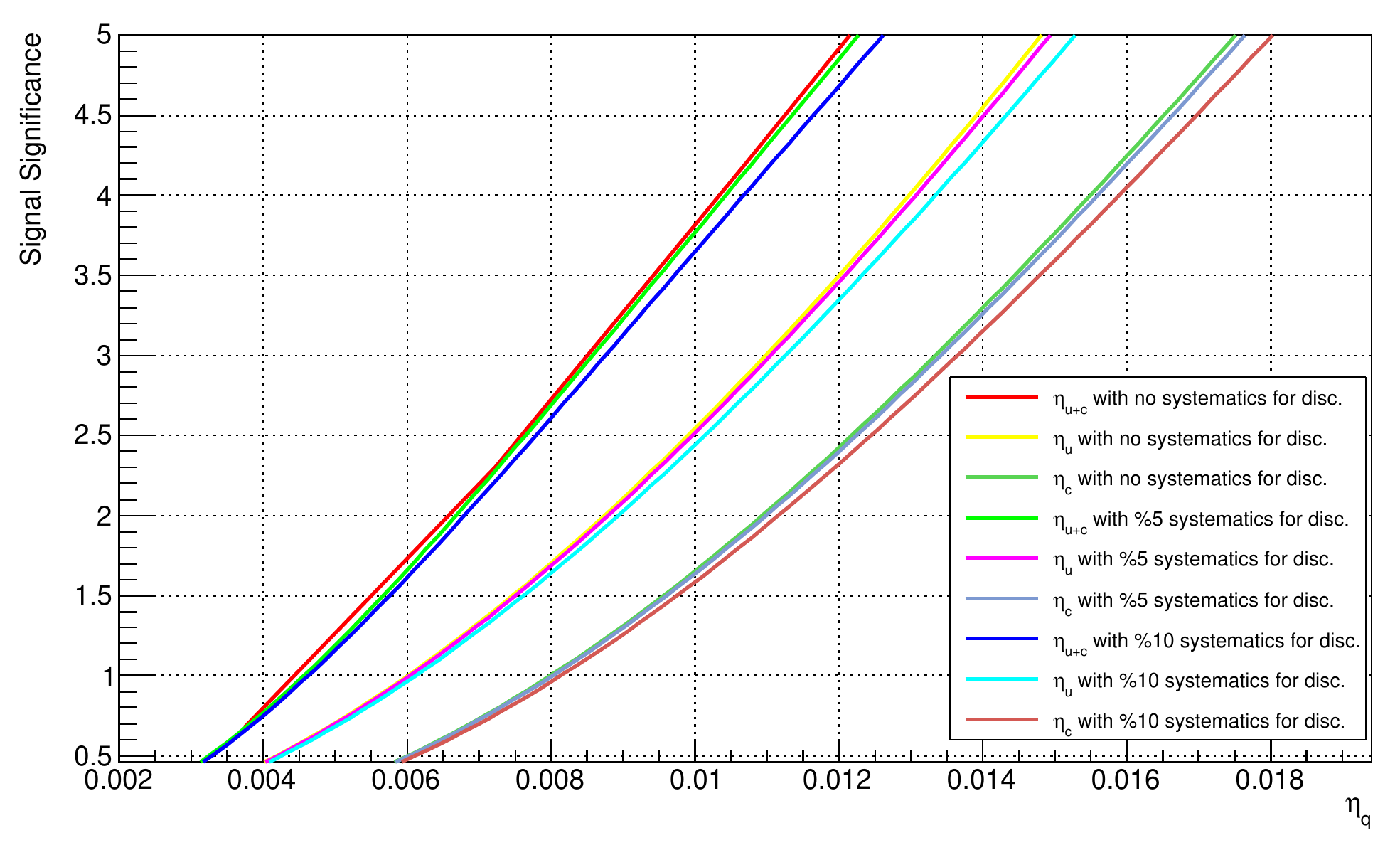}

\caption{Signal significance ($SS$) versus $\eta_{q}$ coupling parameter
for three different scenarios at $3\ \mathrm{ab^{-1}}$ luminosity
for discovery (disc) case.\label{fig:Signal-significance-()}}
\end{figure}

\begin{figure}[h]
\includegraphics[scale=0.44]{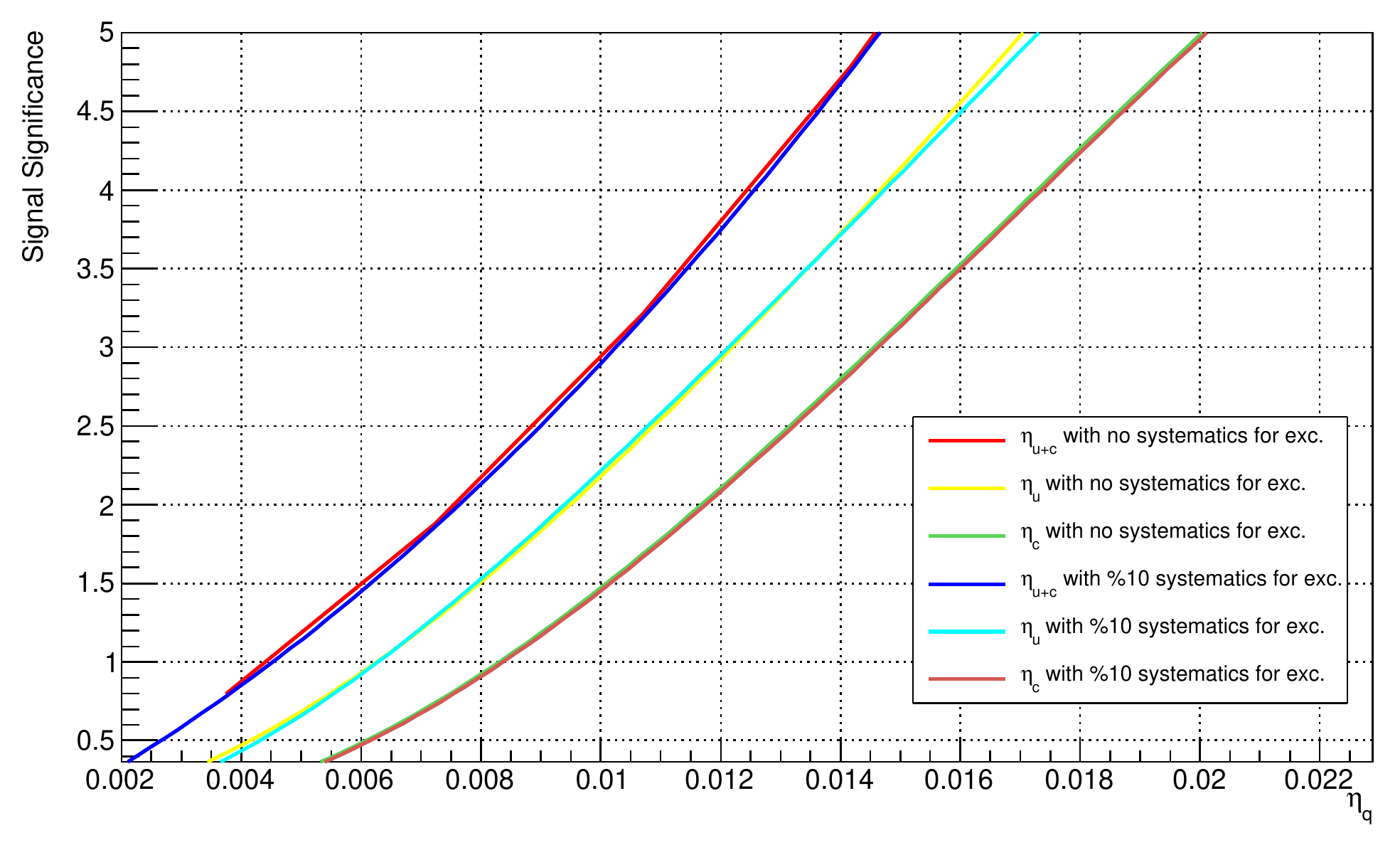}
\raggedright{}\caption{Signal significance ($SS$) versus $\eta_{q}$ coupling parameter
for three different scenarios at $30\ \mathrm{ab^{-1}}$ luminosity
for exclusion (exc) case.\label{fig:Signal-significance-()-123}}
\end{figure}

\begin{figure}[h]
\includegraphics[scale=0.44]{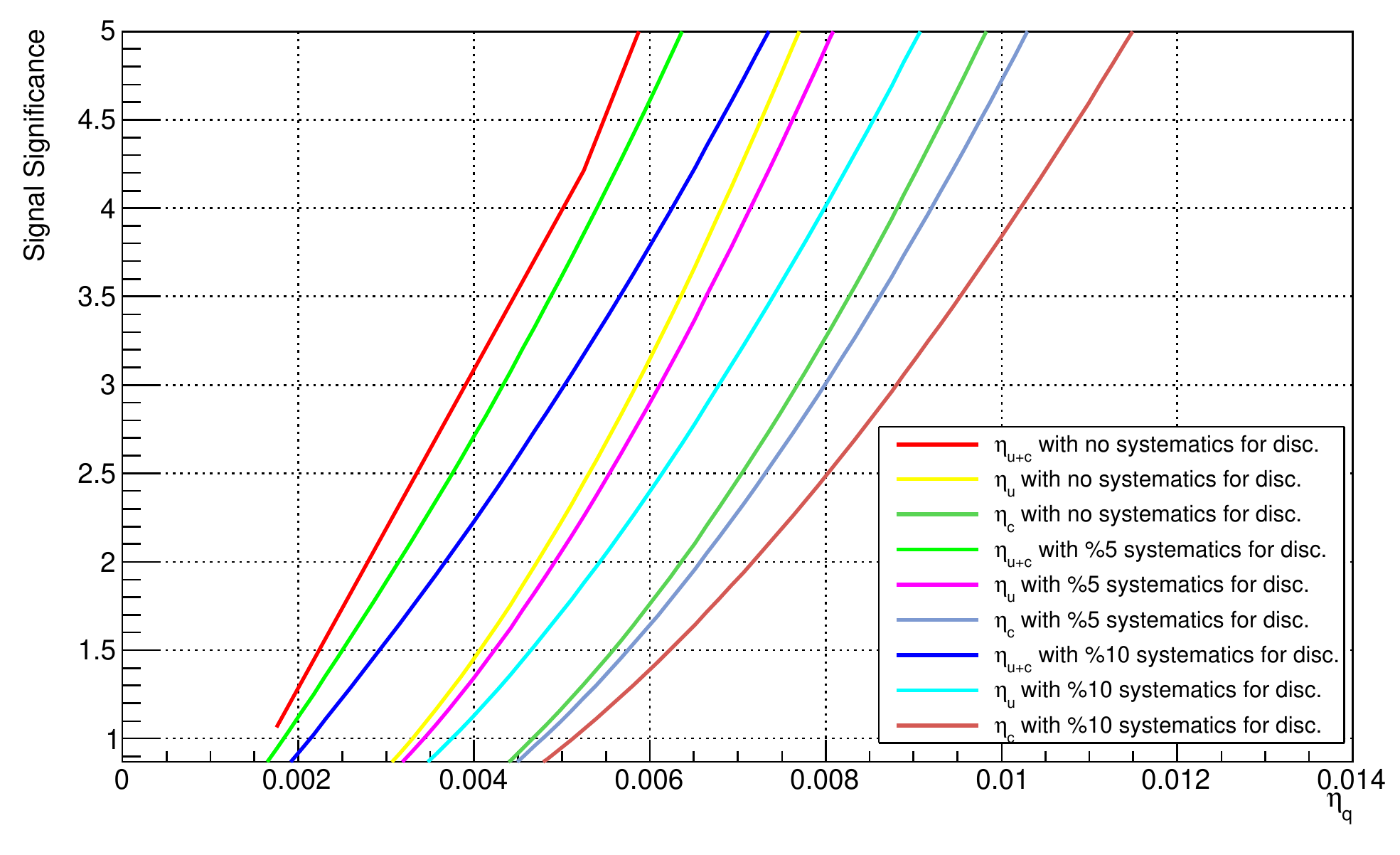}

\caption{Signal significance ($SS$) versus $\eta_{q}$ coupling parameter
for three different scenarios at $30\ \mathrm{ab^{-1}}$ luminosity
for discovery (disc) case.\label{fig:Signal-significance-versus}}
\end{figure}

\begin{figure}[h]
\includegraphics[scale=0.44]{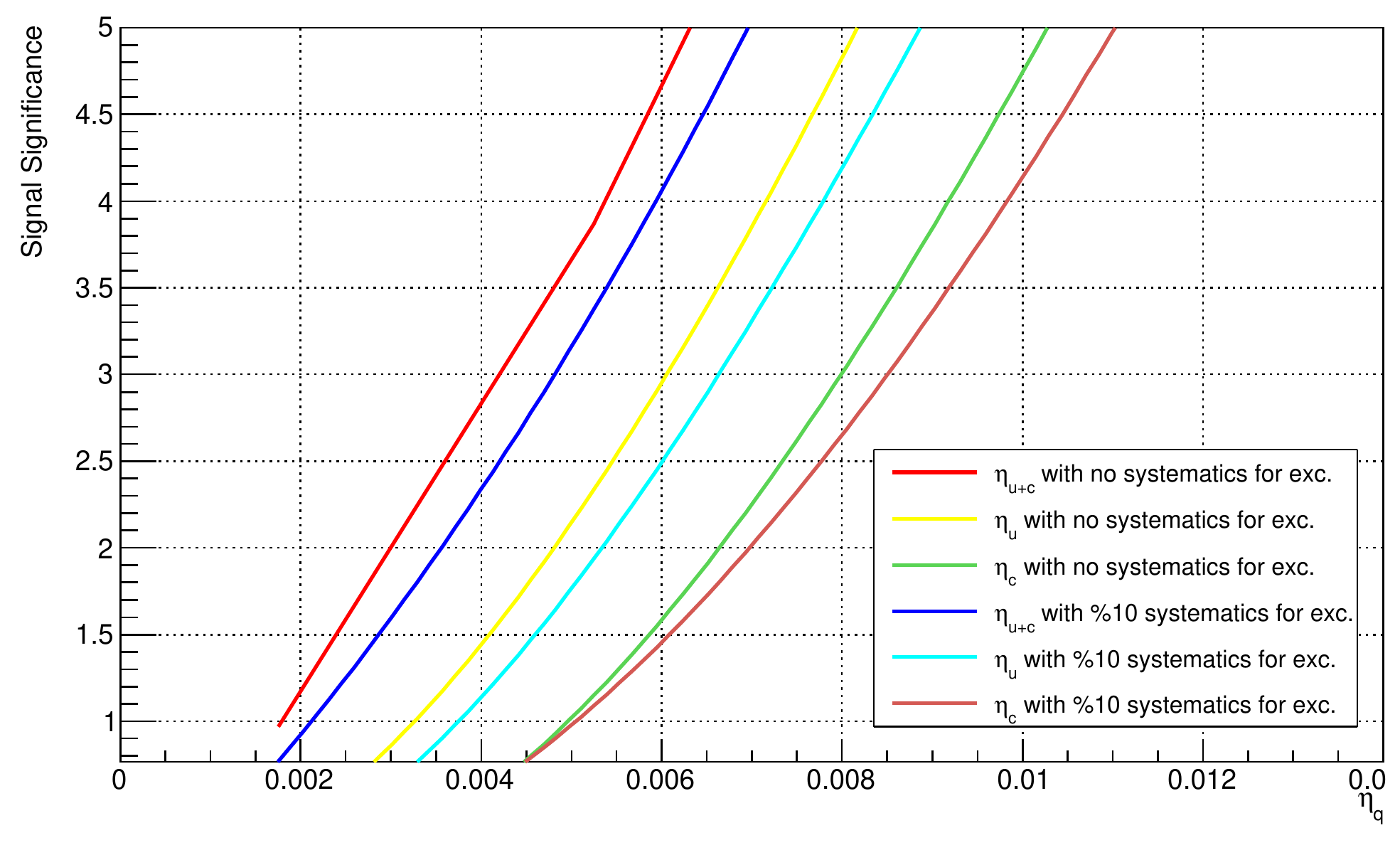}

\caption{Signal significance ($SS$) versus $\eta_{q}$ coupling parameter
for three different scenarios at $30\ \mathrm{ab^{-1}}$ luminosity
for exclusion (exc) case.\label{fig:Signal-significance-()-124}}
\end{figure}

In Table \ref{tab:Upper-limits-on-1}, \ref{tab:Upper-limits-on-2},
\ref{tab:Limitations-on-}, \ref{tab:Upper-limits-on} we summarize
our results with the discovery and exclusion significance according
to $SS_{\mathrm{disc}}$ relations given at Eq. \ref{eq: 4}, \ref{eq:3};
and in Fig. \ref{fig:Signal-significance-()}, \ref{fig:Signal-significance-versus}
we give fit graphs for signal significance for 3 ab$^{-1}$ and 30
ab$^{-1}$ respectively. Similar results have been shown in Fig. \ref{fig:Signal-significance-()-123}
and \ref{fig:Signal-significance-()-124} using $SS_{\mathrm{exc}}$
relations both namely Eq. \ref{eq:5}, \ref{eq:6} which are given
for 3 ab$^{-1}$ and 30 ab$^{-1}$ respectively.
\begin{table}[h]
\caption{Upper limits on $\eta_{q}$ parameter and corresponding branching
ratios at $3\ \mathrm{ab}^{-1}$ luminosity with no systematics.\label{tab:Upper-limits-on-1}}

\begin{raggedright}
\begin{tabular}{|c|c|c|}
\hline 
Scenario & \multicolumn{2}{c|}{$SS_{\mathrm{disc}}\geq2$}\tabularnewline
\hline 
\hline 
 & $\eta_{q}$ & Branching Ratio\tabularnewline
\hline 
$\eta_{u}=\eta_{c}$ & 0.0065 & $1.10\times10^{-5}$\tabularnewline
\hline 
Only $\eta_{u}$ & 0.0088 & $2.02\times10^{-5}$\tabularnewline
\hline 
Only $\eta_{c}$ & 0.0109 & $3.10\times10^{-5}$\tabularnewline
\hline 
\end{tabular}
\par\end{raggedright}
\begin{raggedright}
\begin{tabular}{|c|c|c|}
\hline 
Scenario & \multicolumn{2}{c|}{$SS_{\mathrm{disc}}\geq3$}\tabularnewline
\hline 
\hline 
 & $\eta_{q}$ & Branching Ratio\tabularnewline
\hline 
$\eta_{u}=\eta_{c}$ & 0.0086 & $1.93\times10^{-5}$\tabularnewline
\hline 
Only $\eta_{u}$ & 0.011 & $3.16\times10^{-5}$\tabularnewline
\hline 
Only $\eta_{c}$ & 0.0134 & $4.68\times10^{-5}$\tabularnewline
\hline 
\end{tabular}
\par\end{raggedright}
\raggedright{}%
\begin{tabular}{|c|c|c|}
\hline 
Scenario & \multicolumn{2}{c|}{$SS_{\mathrm{disc}}\geq5$}\tabularnewline
\hline 
\hline 
 & $\eta_{q}$ & Branching Ratio\tabularnewline
\hline 
$\eta_{u}=\eta_{c}$ & 0.0120$\ $ & $3.76\times10^{-5}$\tabularnewline
\hline 
Only $\eta_{u}$ & 0.0148$\ $ & $5.71\times10^{-5}$\tabularnewline
\hline 
Only $\eta_{c}$ & 0.0175$\ $ & $7.99\times10^{-5}$\tabularnewline
\hline 
\end{tabular}
\end{table}
\begin{table}[h]
\caption{Upper limits on $\eta_{q}$ parameter and corresponding branching
ratios at $3\ \mathrm{ab}^{-1}$ luminosity with \%10 systematics.\label{tab:Upper-limits-on-2}}

\begin{raggedright}
\begin{tabular}{|c|c|c|}
\hline 
Scenario & \multicolumn{2}{c|}{$SS_{\mathrm{disc}}\geq2$}\tabularnewline
\hline 
\hline 
 & $\eta_{q}$ & Branching Ratio\tabularnewline
\hline 
$\eta_{u}=\eta_{c}$ & 0.0068 & $1.21\times10^{-5}$\tabularnewline
\hline 
Only $\eta_{u}$ & 0.009 & $2.11\times10^{-5}$\tabularnewline
\hline 
Only $\eta_{c}$ & 0.0112 & $3.27\times10^{-5}$\tabularnewline
\hline 
\end{tabular}
\par\end{raggedright}
\begin{raggedright}
\begin{tabular}{|c|c|c|}
\hline 
Scenario & \multicolumn{2}{c|}{$SS_{\mathrm{disc}}\geq3$}\tabularnewline
\hline 
\hline 
 & $\eta_{q}$ & Branching Ratio\tabularnewline
\hline 
$\eta_{u}=\eta_{c}$ & 0.0088 & $2.02\times10^{-5}$\tabularnewline
\hline 
Only $\eta_{u}$ & 0.0112$\ $ & $3.27\times10^{-5}$\tabularnewline
\hline 
Only $\eta_{c}$ & 0.0136$\ $ & $4.82\times10^{-5}$\tabularnewline
\hline 
\end{tabular}
\par\end{raggedright}
\raggedright{}%
\begin{tabular}{|c|c|c|}
\hline 
Scenario & \multicolumn{2}{c|}{$SS_{\mathrm{disc}}\geq5$}\tabularnewline
\hline 
\hline 
 & $\eta_{q}$ & Branching Ratio\tabularnewline
\hline 
$\eta_{u}=\eta_{c}$ & 0.0126$\ $ & $4.14\times10^{-5}$\tabularnewline
\hline 
Only $\eta_{u}$ & 0.0152$\ $ & $6.02\times10^{-5}$\tabularnewline
\hline 
Only $\eta_{c}$ & 0.018$\ $ & $8.45\times10^{-5}$\tabularnewline
\hline 
\end{tabular}
\end{table}
\begin{table}[h]
\begin{raggedright}
\caption{Upper limits on $\eta_{q}$ parameter and corresponding branching
ratios at $30\ \mathrm{ab}^{-1}$ luminosity with no systematics.\label{tab:Limitations-on-}}
\begin{tabular}{|c|c|c|}
\hline 
Scenario & \multicolumn{2}{c|}{$SS_{\mathrm{disc}}\geq2$}\tabularnewline
\hline 
\hline 
 & $\eta_{q}$ & Branching Ratio\tabularnewline
\hline 
$\eta_{u}=\eta_{c}$ & 0.0027 & $1.90\times10^{-6}$\tabularnewline
\hline 
Only $\eta_{u}$ & 0.0046 & $5.52\times10^{-6}$\tabularnewline
\hline 
Only $\eta_{c}$ & 0.0064 & $1.07\times10^{-5}$\tabularnewline
\hline 
\end{tabular}
\par\end{raggedright}
\begin{raggedright}
\begin{tabular}{|c|c|c|}
\hline 
Scenario & \multicolumn{2}{c|}{$SS_{\mathrm{disc}}\geq3$}\tabularnewline
\hline 
\hline 
 & $\eta_{q}$ & Branching Ratio\tabularnewline
\hline 
$\eta_{u}=\eta_{c}$ & 0.0039 & $3.99\times10^{-6}$\tabularnewline
\hline 
Only $\eta_{u}$ & 0.0059 & $9.08\times10^{-6}$\tabularnewline
\hline 
Only $\eta_{c}$ & 0.0077 & $1.55\times10^{-5}$\tabularnewline
\hline 
\end{tabular}
\par\end{raggedright}
\raggedright{}%
\begin{tabular}{|c|c|c|}
\hline 
Scenario & \multicolumn{2}{c|}{$SS_{\mathrm{disc}}\geq5$}\tabularnewline
\hline 
\hline 
 & $\eta_{q}$ & Branching Ratio\tabularnewline
\hline 
$\eta_{u}=\eta_{c}$ & 0.0059 & $9.08\times10^{-6}$\tabularnewline
\hline 
Only $\eta_{u}$ & 0.0077 & $1.55\times10^{-5}$\tabularnewline
\hline 
Only $\eta_{c}$ & 0.0097 & $2.45\times10^{-5}$\tabularnewline
\hline 
\end{tabular}
\end{table}
\begin{table}[h]
\caption{Upper limits on $\eta_{q}$ parameter and corresponding branching
ratios at $30\ \mathrm{ab}^{-1}$ luminosity with \%10 systematics.\label{tab:Upper-limits-on}}

\begin{raggedright}
\begin{tabular}{|c|c|c|}
\hline 
Scenario & \multicolumn{2}{c|}{$SS_{\mathrm{disc}}\geq2$}\tabularnewline
\hline 
\hline 
 & $\eta_{q}$ & Branching Ratio\tabularnewline
\hline 
$\eta_{u}=\eta_{c}$ & 0.0036 & $3.38\times10^{-6}$\tabularnewline
\hline 
Only $\eta_{u}$ & 0.0054 & $7.60\times10^{-6}$\tabularnewline
\hline 
Only $\eta_{c}$ & 0.0072 & $1.35\times10^{-5}$\tabularnewline
\hline 
\end{tabular}
\par\end{raggedright}
\begin{raggedright}
\begin{tabular}{|c|c|c|}
\hline 
Scenario & \multicolumn{2}{c|}{$SS_{\mathrm{disc}}\geq3$}\tabularnewline
\hline 
\hline 
 & $\eta_{q}$ & Branching Ratio\tabularnewline
\hline 
$\eta_{u}=\eta_{c}$ & 0.0050 & $6.52\times10^{-6}$\tabularnewline
\hline 
Only $\eta_{u}$ & 0.0068 & $1.21\times10^{-5}$\tabularnewline
\hline 
Only $\eta_{c}$ & 0.0088 & $2.02\times10^{-5}$\tabularnewline
\hline 
\end{tabular}
\par\end{raggedright}
\raggedright{}%
\begin{tabular}{|c|c|c|}
\hline 
Scenario & \multicolumn{2}{c|}{$SS_{\mathrm{disc}}\geq5$}\tabularnewline
\hline 
\hline 
 & $\eta_{q}$ & Branching Ratio\tabularnewline
\hline 
$\eta_{u}=\eta_{c}$ & 0.0074 & $1.43\times10^{-5}$\tabularnewline
\hline 
Only $\eta_{u}$ & 0.0090 & $2.11\times10^{-5}$\tabularnewline
\hline 
Only $\eta_{c}$ & 0.012 & $3.76\times10^{-5}$\tabularnewline
\hline 
\end{tabular}
\end{table}
 

As clearly seen from this tables scenario $\eta_{u+c}$ gives better
results for $SS_{\mathrm{disc}}$ (and also for $SS_{\mathrm{exc}}$)
as expected originated from its larger PDF values. For this scenario,
the values mentioned at CDR of FCC-hh \citep{key-7-2} reached and
this scenario is the prospect future restrictor. Note that, the other
scenarios follows $\eta_{u+c}$ case closely, hence the found limits
change accordingly. Normally one may give no special importance to
the other scenarios unless an additional mechanism blocks the transtion
from both quarks to top. However, as we are dealing with an effective
theory, such a mechanism is quite reasonable (yet needed to be observe),
so we would like to cover this cases for any possible FCNC scenario.
For $\eta_{u}$ and $\eta_{c}$ scenarios, our results for limitations
on coupling constants are still better than known LHC limits and will
give the first implications of FCNC interactions with hinting an additional
quark transition preserving mechanism. The effects of systematics
gives no values so much different than the ideal case due to highly
reduced background. This is the power of our analysis that role of
systematics lowered in this sense. Moreover, as we know that the backgrounds
have large number of cross-section so, even \%5 of uncertainty is
disastereous, but tamed for our analysis. The exclusion regions are
more or less same for $SS_{\mathrm{disc}}$ and $SS_{\mathrm{exc}}$,
hence we can say compatible with each in this sense. To be specific
we can push exclusion limits at best to arround 0.00265 which decides
the fate of FCNC type interactions. At $3\ \mathrm{ab}^{-1}$ luminosity,
and as a worst case scenario with \%10 systematics included even only
$c$ case, coupling constants below the recent limits. For more luminosity
results will be enhanced obviously. In literature our results are
compatible with other phenomenological results \citep{key-10-1,key-10-2,key-10-3,key-10-4}
(Bear in mind that our effective Lagrangian is so restrictive in general,
thus limits can be imporeved directly by 4 times when we consider
coupling constants).

In this study we have tried to estimate searches using the channel
which includes single production of top quark (decay leptonically)
with a Higgs boson (decay a bottom pair) via FCNC interaction at FCC-hh.
This collider offers large number of events at high energies compared
to its predeccesor and enhance the potential of new physics searches
greatly. Former searches done by both the CMS and the ATLAS collabarations
extracts the current limitations of FCNC interactions, \citep{key-6-1,key-6-2,key-6-3,key-6-4}
nevertheless our research at FCC-hh has potentially decide the limits
on coupling constants. Furthermore, first implications of MSSM and
2HDM (FC) can be tested.

This rare process suffers high backgrounds even more in higher energies.
To overcome this difficulty one needs to elaborate this type of analysis
much more carefully. As explained in detail in analysis section, we
regard this features of processes when distinguishing signal from
background. We found accessible limits for coupling constants which
have been considerably enhanced, furthermore they are even better
compared the results given in literaure. Besides, we see the value
of the coupling constants become very small which hints the interaction
is so weak. 

In conclusion, it seems that $pp\rightarrow th(j)$ channel plays
an important role for future coliders. The significance of observation
or discovery for the FCNC interactions at this channel is fairly high
if these kind of interactions are really exist in nature. Our results
show that potential discovery or exclusion limits on coupling constants
for FCNC interactions can be set arround 0.0059 (which corresponds
to $BR(\eta_{u+c})_{\mathrm{disc}}=9.08\times10^{-6}$) or 0.0027
corresponding to $BR(\eta_{u+c})_{\mathrm{exc}}=2.78\times10^{-6}$,
respectively. This will extract nearly full potential of FCC-hh collider
for $tqH$ FCNC interactions.
\begin{acknowledgments}
No funding was received for this research.

The authors are grateful to Ulku Ulusoy for a careful reading of the
manuscript. We wish to acknowledge the support of the AUHEP group,
offering suggestions and encouragement. The numerical calculations
reported in this paper were partially performed at TUBITAK ULAKBIM,
High Performance and Grid Computing Center (TRUBA resources).

\newpage{}
\end{acknowledgments}

\end{document}